\documentclass[twocolumn]{aastex631}

\usepackage{CJK}
\usepackage{pifont}
\usepackage{xcolor}
\usepackage{booktabs}
\definecolor{GoodMark}{RGB}{0,90,180}
\definecolor{BadMark}{RGB}{200,90,0}  
\begin{document}

\title{3D Kinematic Reconstruction of the Crab Nebula That Includes the Northern Ejecta `Jet'}

\begin{CJK*}{UTF8}{gbsn}

\author[0009-0003-0122-9472]{Z.\ Ding (丁子为)}
\affiliation{Department of Physics and Astronomy, Purdue University, 525 Northwestern Avenue, West Lafayette, IN 47907, USA}

\author[0000-0002-0763-3885]{D.\ Milisavljevic}
\affiliation{Department of Physics and Astronomy, Purdue University, 525 Northwestern Avenue, West Lafayette, IN 47907, USA}

\author{T.\ Martin}
\affiliation{D\'epartement de physique, de g\'enie physique et d'optique, Universit\'e Laval, Qu\'ebec, QC G1V 0A6, Canada}

\author[0000-0001-7380-3144]{T.\ Temim}
\affiliation{Princeton University, 4 Ivy Ln, Princeton, NJ 08544, USA}

\author[0000-0002-7868-1622]{J.\ C.\ Raymond}
\affiliation{Center for Astrophysics $\vert$ Harvard \& Smithsonian, 60 Garden Street, Cambridge, MA 02138, USA}

\author[0000-0001-9484-1262]{Soham Mandal}
\affiliation{Department of Astronomy, University of Virginia, 530 McCormick Road, Charlottesville, VA 22904, USA}
\affiliation{Virginia Institute for Theoretical Astronomy, University of Virginia, Charlottesville, VA 22904, USA}

\author[0000-0003-1278-2591]{L.\ Drissen}
\affiliation{D\'epartement de physique, de g\'enie physique et d'optique, Universit\'e Laval, Qu\'ebec, QC G1V 0A6, Canada}

\begin{abstract}
We present new detailed three-dimensional kinematic reconstructions of the Crab Nebula created from hyperspectral cubes obtained with the SITELLE instrument mounted on the Canada-France-Hawaii Telescope. Our data cubes span a wavelength range from 3600\r{A} to 7000\r{A} covering major emission lines including [\ion{O}{2}] $\lambda$$\lambda$3726, 3729, $H\beta$, [\ion{O}{3}] $\lambda$$\lambda$4959, 5007, [\ion{N}{2}] $\lambda$5755, \ion{He}{1} $\lambda$5876, [\ion{N}{2}] $\lambda\lambda$6548, 6584, [\ion{S}{2}] $\lambda\lambda$6717, 6731, and $H\alpha$. The field of view encompasses the ``chimney'' or ``jet," which is a 45-arcsec wide funnel-shaped structure that stretches 100 arcsec off of the northern limb. 
Our 3D reconstructions confirm and geometrically resolve a cavity at the jet's base suggested by earlier kinematic studies, establishing a direct physical connection between the filamentary network and the jet funnel. The morphology and kinematics indicate that the early pulsar wind nebula (PWN) played a central role in forming the jet. Several formation scenarios that are not necessarily mutually exclusive remain viable, including a bipolar outflow shaped by a circumstellar disk, a breach or underdensity in the ejecta shell, and a pre-existing progenitor mass loss trail acting as a low density channel. Collectively, these scenarios exhibit differing capacities to account for the jet's pronounced collimation, the absence of a southern counterpart, and its near-ballistic motion. Discriminating among them will require fully three-dimensional hydrodynamic simulations that trace the remnant's evolution from the progenitor phase through late-time PWN expansion.

\end{abstract}


\section{Introduction} \label{sec:intro}
The Crab Nebula, the remnant of the core-collapse supernova SN\,1054 \citep{lundmark1921}, consists of three components: the Crab pulsar at the center, the Crab synchrotron nebula filled with relativistic plasma, and a complex of ejecta filaments toward the outer edge of the synchrotron nebula (see reviews in \citealt{DF85} and \citealt{hester2008crab}, as well as recent high-resolution imaging studies by \citealt{Blair2026}). These filaments of ejecta are also referred as the ``[\ion{O}{3}] skin'' because of their optimal visibility in [\ion{O}{3}] $\lambda$$\lambda$4959, 5007 line emission \citep{sankrit1997}. 

The [\ion{O}{3}] skin layer includes a remarkable filamentary feature, 45 arcsec wide and extending approximately 100 arcsec off the nebula's northern limb, also known as the ``chimney'' or ``jet" \citep{vandenbergh1970,Velusamy1984,Black2015}. This optical structure should not to be confused with the physically smaller X-ray synchrotron jets off of the central pulsar \citep{weisskopf2000,rudie2008}. 

\begin{figure*}[tp]
\centering

\includegraphics[width=0.9\linewidth]{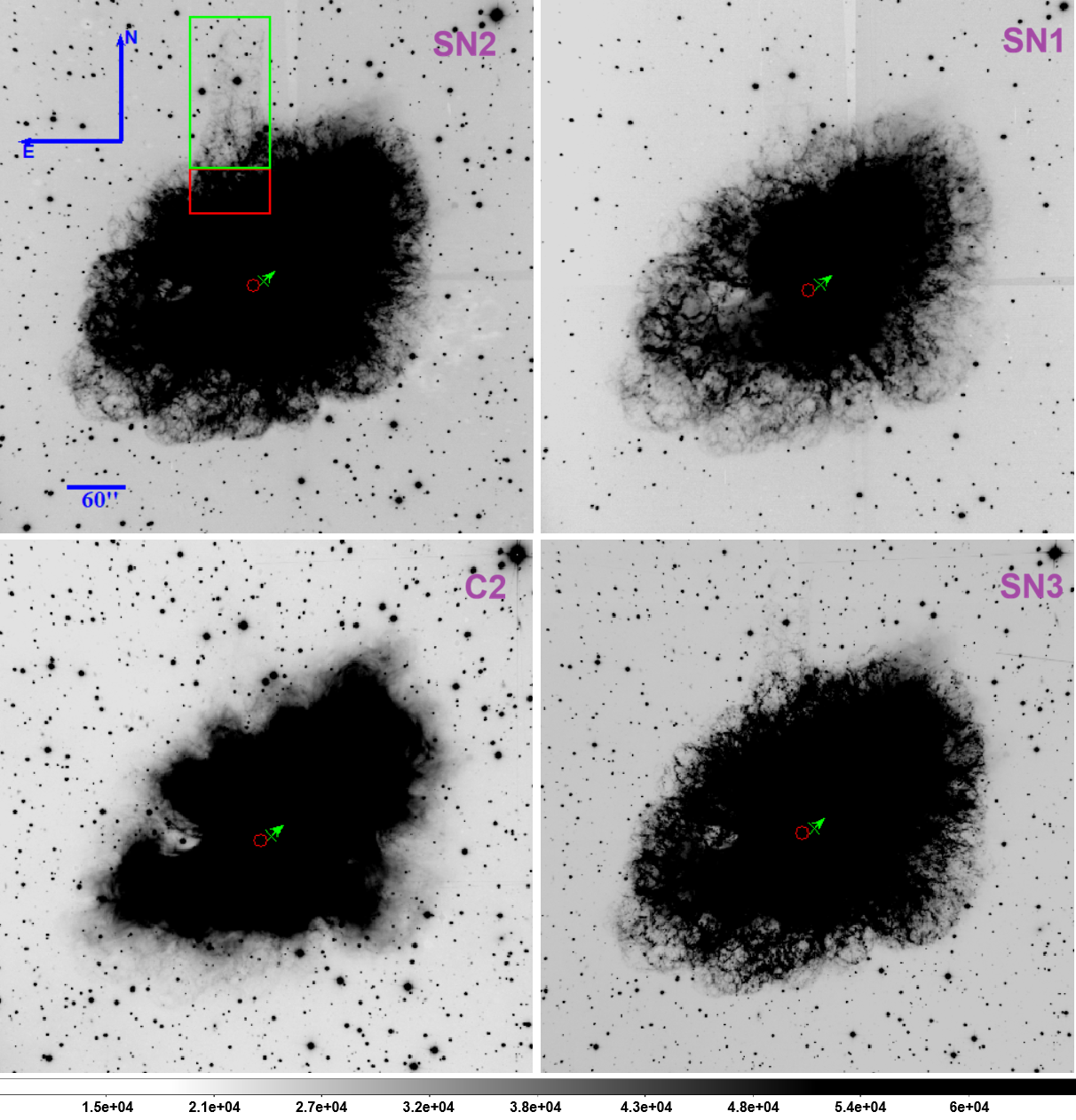}

\captionsetup{font=small}
\caption{Deep frame images extracted from our SITELLE spectral cubes of the Crab Nebula. All images are displayed using the same intensity scale, as indicated by the intensity bar at the bottom of the figure. The green box indicates the jet region and the red box indicates the hole region (see section 4.3). The red circle represents the expansion center \citep{Nugent1998} and the green ``X'' represents the location of the Crab pulsar \citep{Kaplan2008}. An arrow is given to indicate the direction of the Crab pulsar's proper motion as determined by \cite{Roger2006}.}
\label{fig:deepframe}
\end{figure*}

The jet is virtually hollow of [\ion{O}{3}] emission, and elliptical in shape, with radial velocities (line-of-sight) ranging from -190 to +480 km s$^{-1}$ and transverse velocities (plane of the sky) of 1600 to 2650 km s$^{-1}$ from base to top \citep{Black2015}. Other observations show that the jet is slightly tilted into the plane of the sky by $\approx$ 8$^{\circ}$ \citep{shull1984}, with an estimated mass of  $\approx$ 0.003 $M_\odot$ \citep{veron1985}. Continuum emission consistent with synchrotron radiation has been observed at optical and radio wavelengths within the boundaries of the jet walls \citep{Velusamy1984,Woltjer87}.

The origin of the Crab's jet has been the subject of debate for decades, and numerous mechanisms have been proposed to explain its unusual, highly collimated morphology. One early theoretical interpretation proposed that the jet formed as a consequence of magnetic instabilities within the nebula's filamentary shell. In this model, long, narrow channels of relatively ``strong'' ($\approx{10^{-3}}$G) and ``weak'' ($\approx{10^{-4}}$G) magnetic fields become squeezed between adjacent filaments by plasma instabilities.  As these confined magnetic structures expand outward, they naturally produce the jet's elongated, cylindrical morphology \citep{bychkov1975,marcelin1990}.

\begin{deluxetable*}{cccccccc}
\tabletypesize{\footnotesize}
\tablewidth{0pt}
\tablecaption{Data from SITELLE spectral cubes presented in this paper \label{cubes}}
\tablehead{
\colhead{Filter} & 
\colhead{Wavelength} & 
\colhead{Resolution} & 
\colhead{Int. Time (s)} & 
\colhead{FWHM (arcsec)} & 
\colhead{Emission Lines} & 
\colhead{SNR} & 
\colhead{Data Points} 
}
\startdata
SN1 & 365--385\,nm & 4818 & 8892 & $1.33\pm0.05$ & [\ion{O}{2}] $\lambda\lambda$3726, 3729 & 6 & 96736 \\
SN2 & 480--520\,nm & 4853 & 7094 & $1.16\pm0.07$ & [\ion{O}{3}] $\lambda\lambda$4959, 5007, H$\beta$ & 10 & 346629 \\
C2 & 562--625\,nm & 1412 & 15520 & $1.17\pm0.06$ & [\ion{N}{2}] $\lambda$5755, \ion{He}{1} $\lambda$5876 & 5 & 70899 \\
SN3 & 651--685\,nm & 4768 & 9317 & $0.94\pm0.08$ & H$\alpha$, [\ion{N}{2}] $\lambda\lambda$6548,6584, [\ion{S}{2}] $\lambda\lambda$6717,6731 & 10 & 755912 \\
\enddata
\end{deluxetable*}

Many models have invoked the dynamical influence of the Crab's pulsar wind nebula (PWN). In these scenarios, the combined action of the PWN and its magnetic field drive a Rayleigh-Taylor-unstable ``bubble'' that pushes through the ejecta shell to produce the jet. Because this unstable interface does not efficiently form new filaments in the surrounding [\ion{O}{3}] skin, the region above the breakout remains open, giving the appearance that the jet is simply a continuation of the [\ion{O}{3}] emitting shell \citep{chevalier1975,sankrit1997,Smith2013,Blondin2017}. A related idea invokes the influence of an inhomogeneous interstellar medium (ISM). If the Crab Nebula encountered a region of significantly lower density within the ISM, the Rayleigh-Taylor-unstable relativistic cavity could preferentially expand into this weakly confining medium, carving a cylindrical breakthrough that mimics a jet \citep{kundt1983, veron1985,Porth2014}.

A wide range of additional related explanations have also been considered. These include the possibility that the jet traces a mass-loss trail left behind by the red giant progenitor system \citep{blandford1983, Cox1991}, a relativistic pulsar plasma beam driven directly by the pulsar \citep{shull1984, benford1984,michel1985,bietenholz1990}, and the ``shadowed flow" model in which a small interstellar cloud blocks the expansion of part of the filamentary shell, diverting material into a narrow outflow along the jet's axis \citep{morrison1985}.

A final class of interpretations argues that the jet traces the highest-velocity ejecta in the Crab's north-south bipolar expansion \citep{fesen1993,fesen1997,rudie2008}. The remnant's pinched expansion near the east-west band of He-rich filaments \citep{MacAlpine1989}, and the east and west ``bays'' of the synchrotron nebula \citep{Fesen1992}, may reflect circumstellar material that inhibited expansion in this direction \citep{MacAlpine1989,Fesen1992,Smith2013}, potentially directing faster outflow along the N-S axis. In this view, the alignment between the remnant's center of expansion and the well-defined blueshifted near side and redshifted far side of the jet suggests a relationship between the jet and the overall expansion geometry. Moreover, the jet's placement above a large, nearly emission-free opening in the thick outer ejecta shell is consistent with the jet forming the polar extension of this N-S bipolar structure \citep{Black2015}.

\begin{figure*}[tp]
\centering

\begin{subfigure}{0.98\textwidth}
    \centering
    \includegraphics[width=\linewidth,height=0.19\textheight,keepaspectratio]{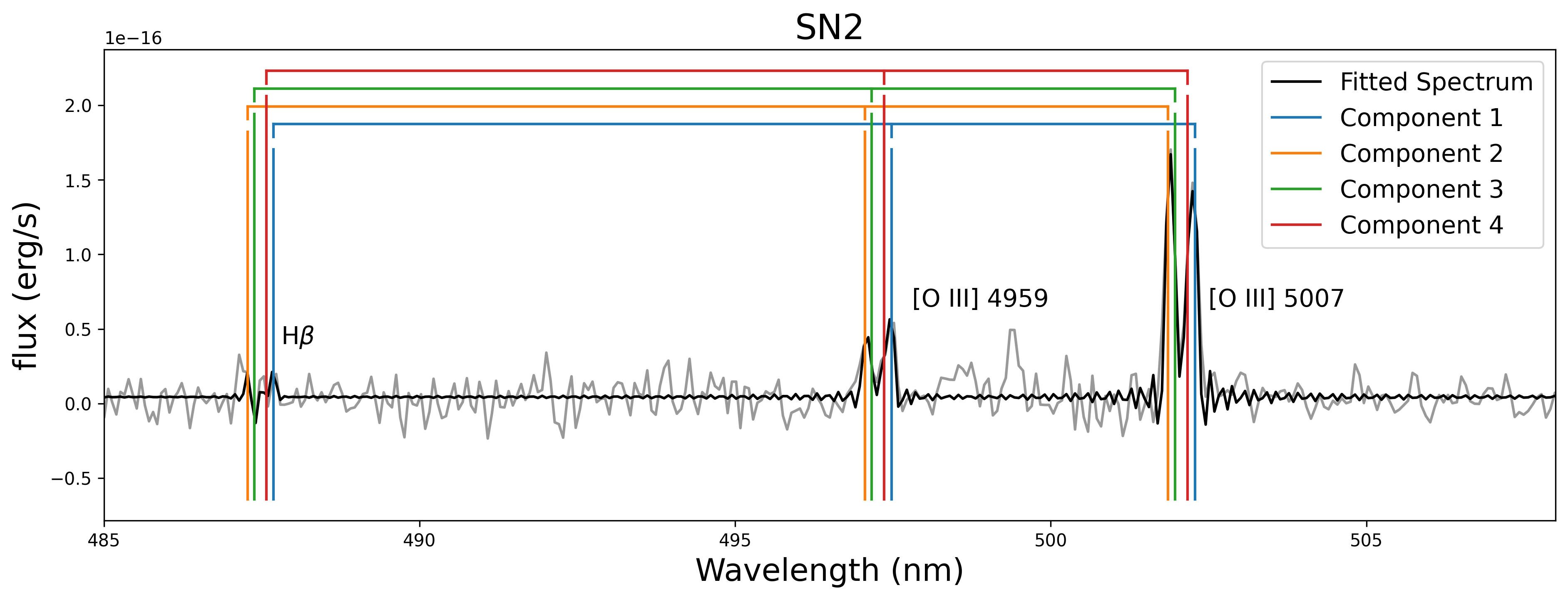}
\end{subfigure}

\vspace{-0.1em}

\begin{subfigure}{0.98\textwidth}
    \centering
    \includegraphics[width=\linewidth,height=0.19\textheight,keepaspectratio]{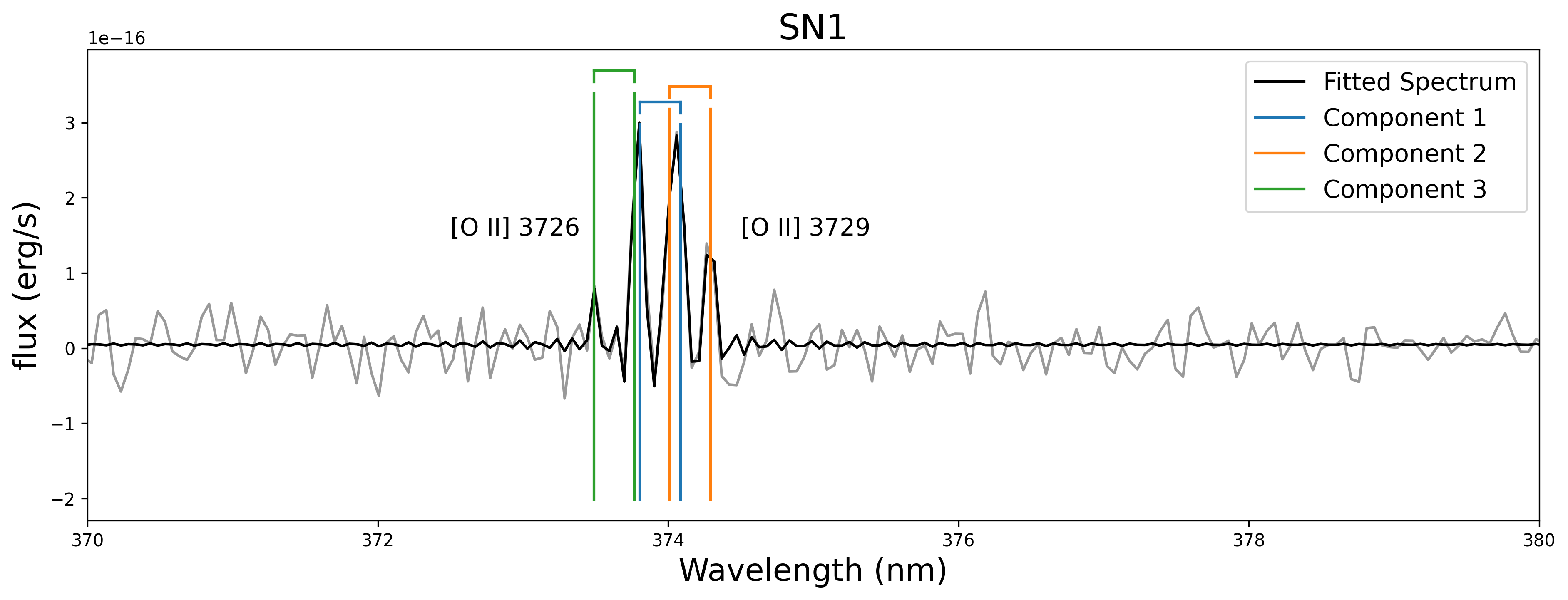}
\end{subfigure}

\vspace{-0.1em}

\begin{subfigure}{0.98\textwidth}
    \centering
    \includegraphics[width=\linewidth,height=0.19\textheight,keepaspectratio]{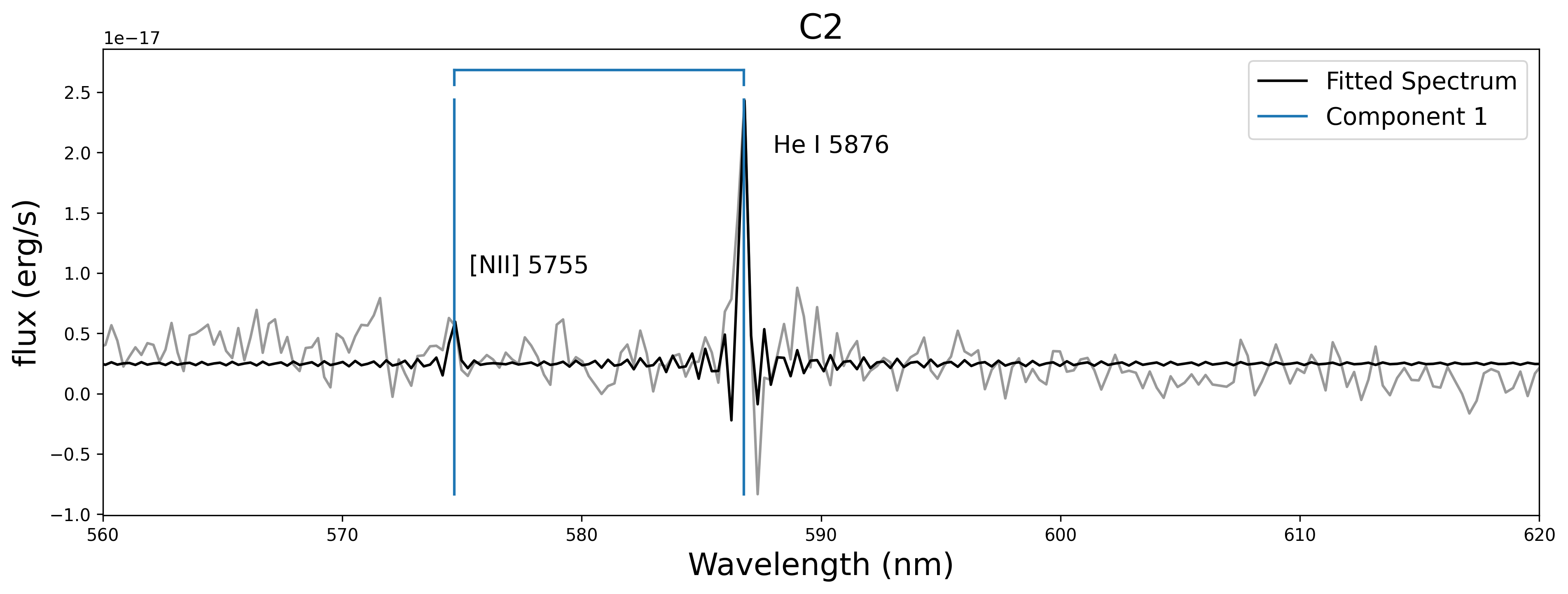}
\end{subfigure}

\vspace{-0.1em}

\begin{subfigure}{0.98\textwidth}
    \centering
    \includegraphics[width=\linewidth,height=0.19\textheight,keepaspectratio]{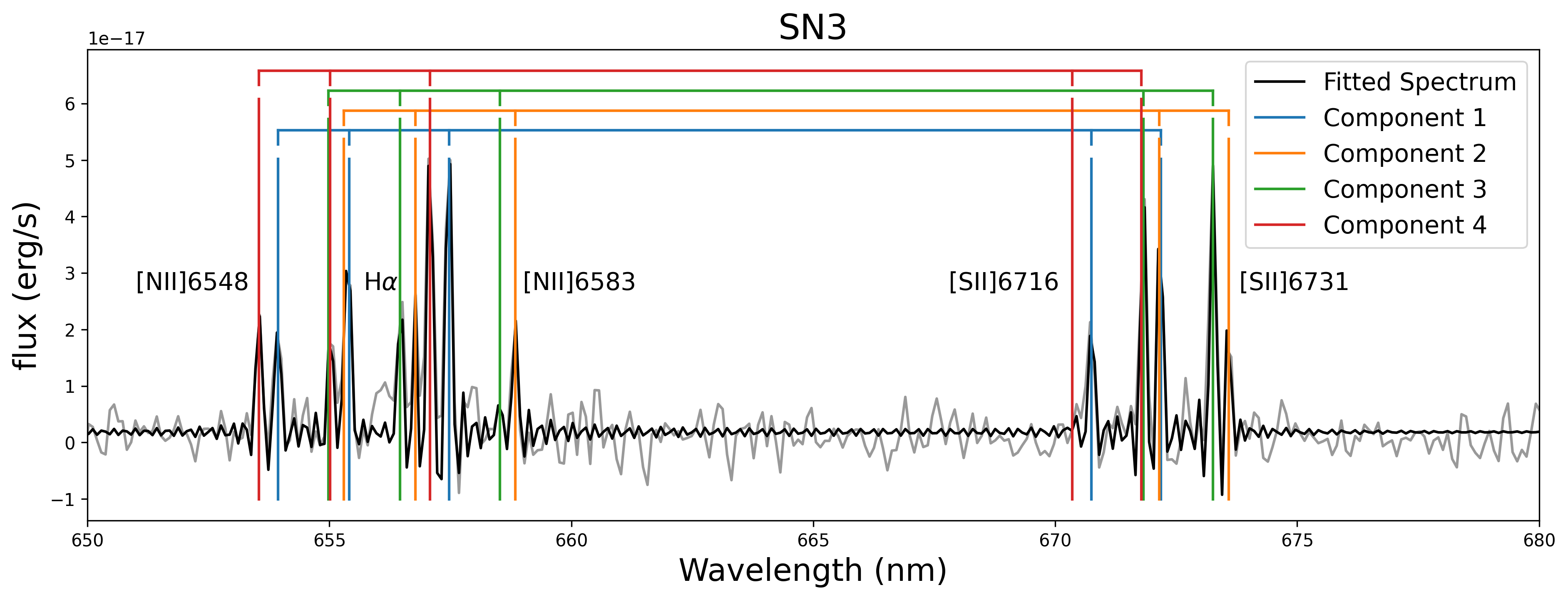}
\end{subfigure}

\captionsetup{font=small}
\caption{Sample spectra demonstrating the quality of our observations and how our algorithm measures emission lines with possible multiple components using cross-correlation. Each spectrum was selected from a different location within the remnant.}
\label{fig:spectrum}
\end{figure*}

In order to have a better understanding of the Crab jet and its relationship with the bulk of the nebula, we present a new three-dimensional Crab Nebula reconstruction of optically-emitting material created from data obtained with SITELLE. This work follows the previous 3D reconstruction of the Crab Nebula in \cite{Martin2021}, which utilized a spectral cube created from the SITELLE observations obtained in 2016. The previous work covered H$\alpha$, [\ion{N}{2}] $\lambda\lambda$6548, 6583, and [\ion{S}{2}] $\lambda\lambda$6717, 6731 line emissions, wh are poor tracers of the jet. The paper is organized as follows.  We describe SITELLE observations in Section 2 and the data reduction process in Section 3. In Section 4, we present the results of the new 3D reconstruction of the Crab Nebula, including details about the jet and newly discovered structures. Section 5 discusses the possible origins of the jet and other newly identified structures. Conclusions are summarized in Section 6. 

\section{Observations} \label{sec:style}

Our data were obtained using the SITELLE imaging Fourier transform spectrometer \citep{Drissen2019} mounted on the Canada-France-Hawaii telescope (CFHT) through two different programs. The first program (ID: 20BC02) observed the Crab Nebula with the SN1, SN2, and C2 filters during the nights of October 13th, November 14th, and November 17th, 2020. The second program (ID: 22BD14) observed the Crab Nebula with SN3 filter during the night of December 27th, 2022. 

SITELLE combines a 2D imaging detector with a Michelson interferometer. Two complementary interferometric data cubes are obtained by recording images, on two 2k×2k CCD detectors, at different positions of the moving mirror inside the Michelson. Fourier transforms are then used to convert these cubes into a single spectral data cube. Spectral resolution is set by the maximum path difference between the two arms of the interferometer, reached by displacing its moving mirror through a series of steps of several hundred nanometers each. The spectral range is selected by using interference filters; SITELLE covers the 350 - 850 nm range with a series of 8 filters, tailored to specific needs. Spatial sampling is 0.32'' per pixel, leading to a field of view of $11' \times 11'$ and over 4 million spectra.

The wavelength range, resolution, emission lines spanned, and integration time for each filter  SN1, SN2, C2, and SN3 is summarized in Table~\ref{cubes}. All data were reduced with the pipeline reduction software ORBS \citep{Martin2012,Martin2015} without any special treatment with respect to the rest of the data obtained using the other filters. 

\section{DATA}

Figure~\ref{fig:deepframe} shows the deep frame images of all filters made from stacking all integrations. For data in each filter, we follow the same processes used to convert SITELLE observations from 2016 into a 3-dimensional Crab Nebula data cube presented in \cite{Martin2021}. 

Given a spectrum at a specific location in the nebula obtained from observations, an automatic algorithm is used to detect and fit the overlapping emission components. The algorithm follows three steps: (i) evaluating the probability expressed as a score of having an emission component at a given velocity; (ii) identifying all individual velocity components above the threshold score along the line of sight (LOS); and (iii) fitting the spectrum with a model that simultaneously includes all detected velocity components \citep{Martin2021}. For each filter, different line ratio constraints were applied to reject obviously incorrect scores during step (i). For example, we reject the component with higher [\ion{O}{3}] $\lambda$4959 score than [\ion{O}{3}] $\lambda$5007. Furthermore, the fitting was performed independently for each filter, with no cross-correlations imposed between filters.

A selection of sample spectra is shown in Figure~\ref{fig:spectrum} to illustrate how the automatic algorithm extracts the different velocity components of the ejecta from specific locations in the nebula with different observation filters.


With all of the velocity components identified by the algorithm for each spectrum, the flux and radial velocity of the ejecta can be determined at every location in the remnant. We then converted the radial velocities to radial distances \textit{z} (LOS) using the age of the remnant, along with the expansion factor $e = 1.160(15)\times{10^{-3}}{\text{yr}^{-1}}$, computed by \citet{Nugent1998}. For the x-y plane, 1$^{\prime\prime}$= $9.696\times{{10}^{-3}}$ pc was used to calculate the x-y distance to the expansion center of the nebula \citep{Martin2021} by assuming a distance of 2 kpc to the remnant \citep{Trimble1973}.

\begin{figure*}
\begin{subfigure}{.5\textwidth}
  \centering
  \includegraphics[width=1\linewidth]{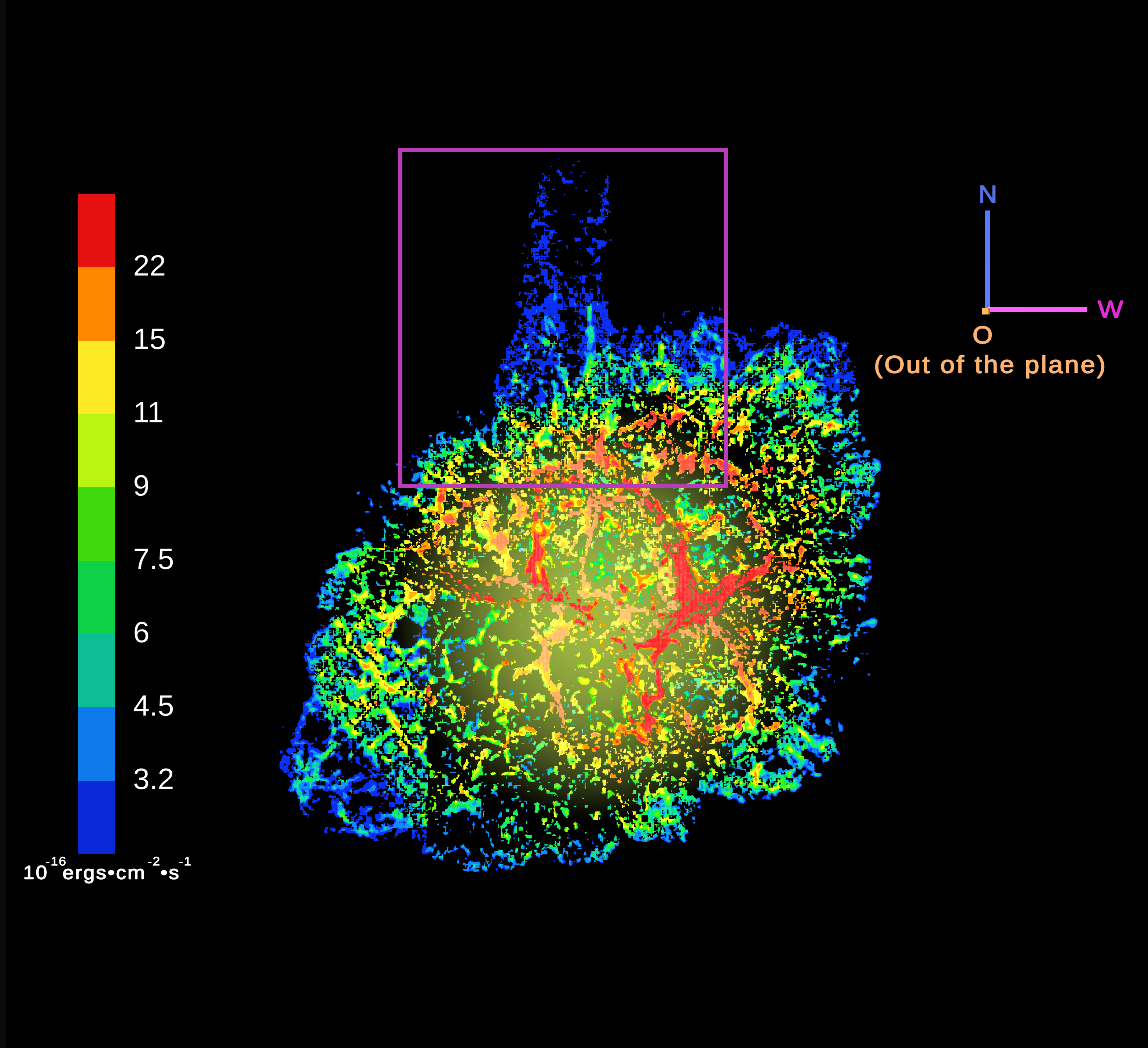} 
\end{subfigure}
\begin{subfigure}{.5\textwidth}
  \centering
  \includegraphics[width=1\linewidth]{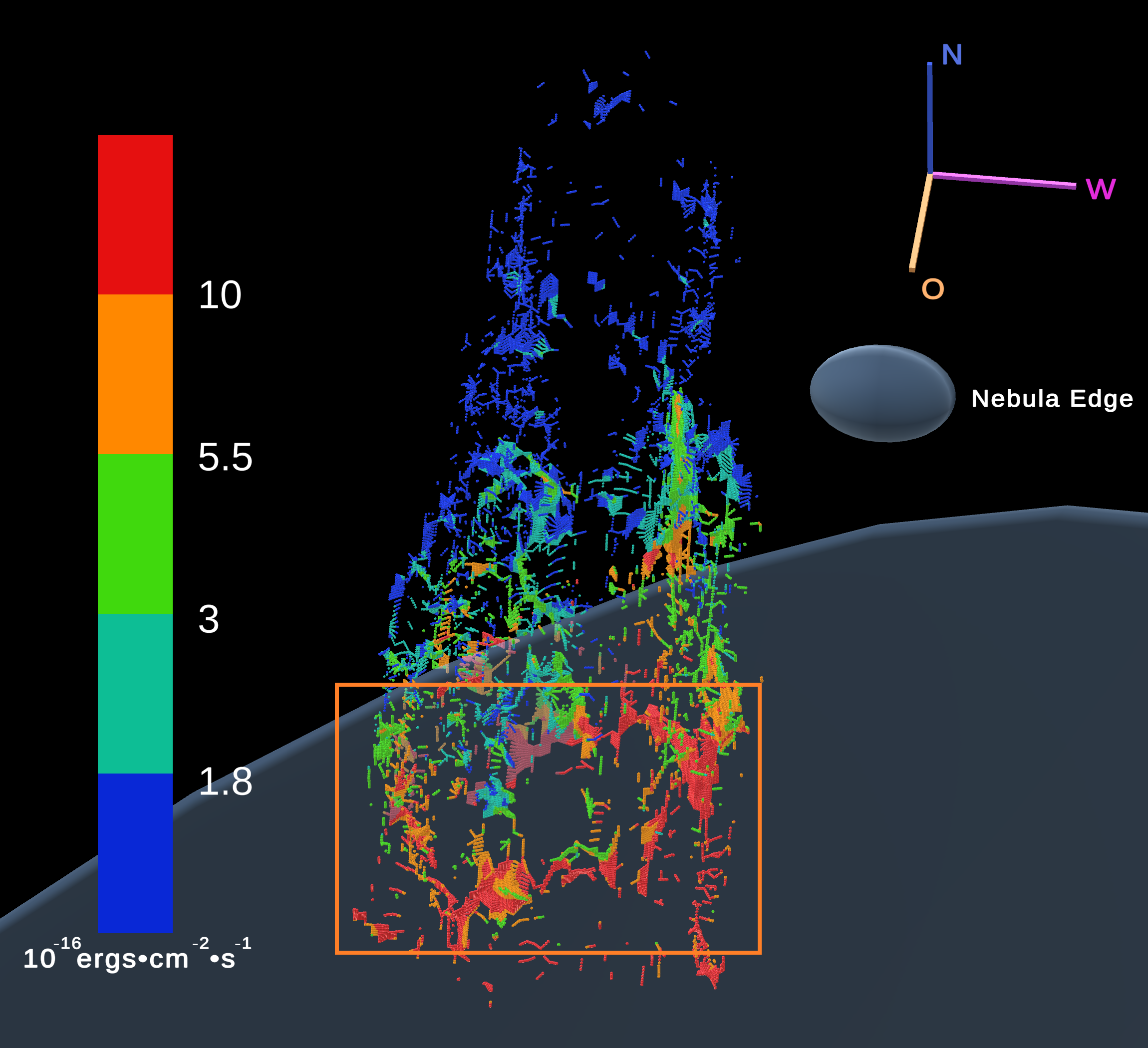}  
\end{subfigure}
\begin{subfigure}{.5\textwidth}
  \centering
  \includegraphics[width=1\linewidth]{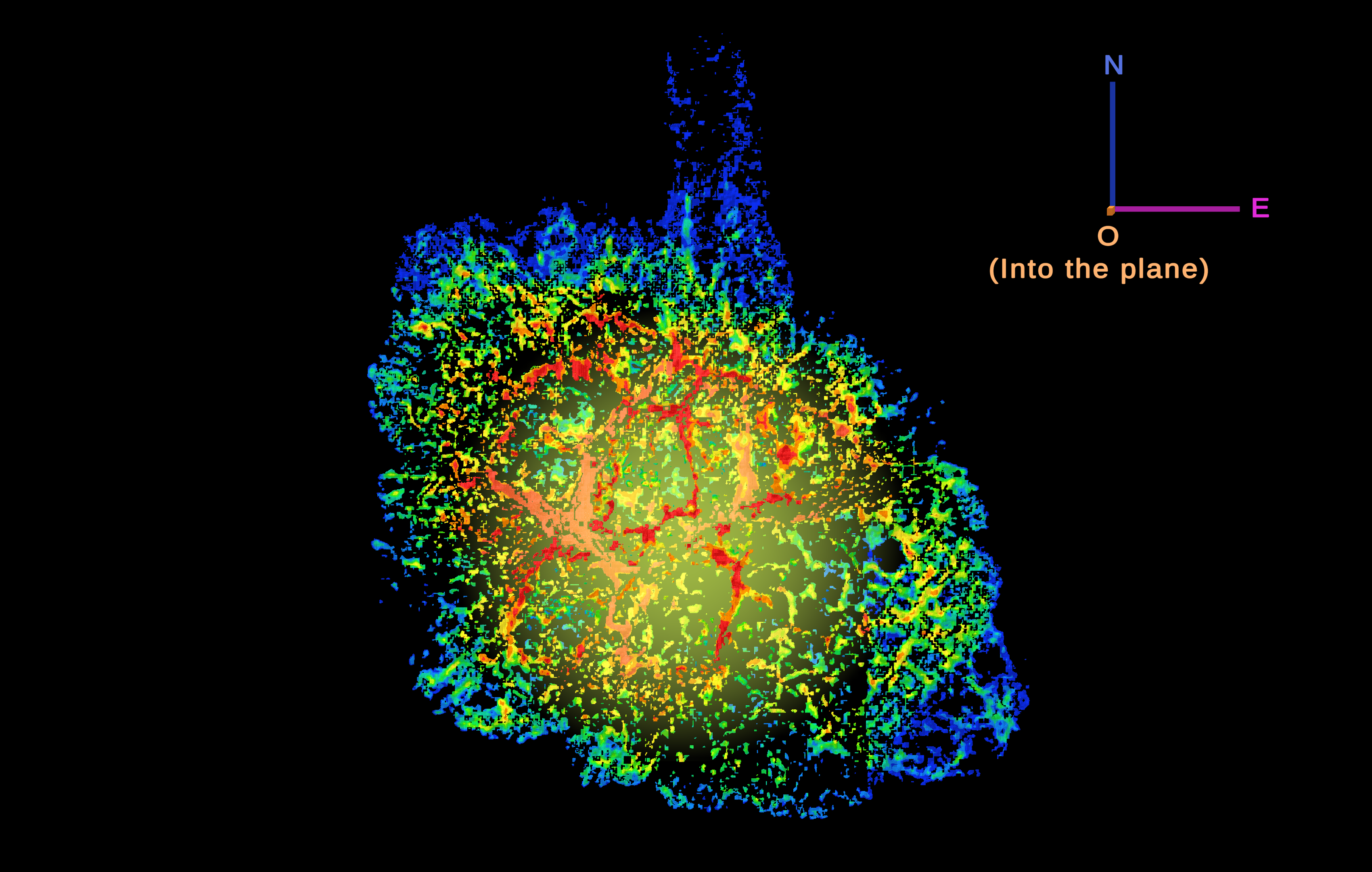} 
\end{subfigure}
\begin{subfigure}{.5\textwidth}
  \centering
  \includegraphics[width=1\linewidth]{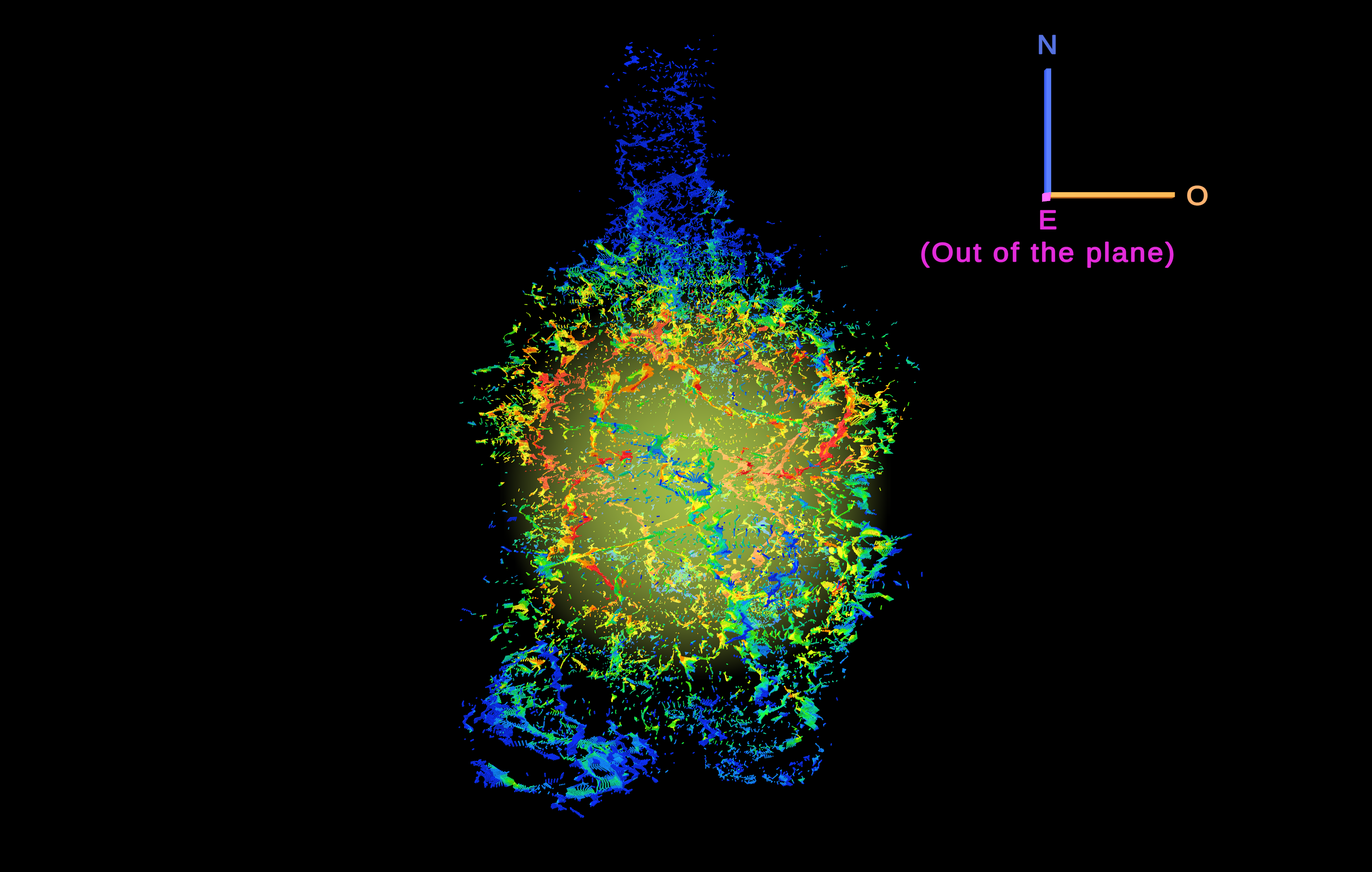} 
\end{subfigure}
\captionsetup{font=small}
\caption{3-dimensional reconstruction of Crab Nebula in [\ion{O}{3}] $\lambda$$\lambda$4959, 5007 emission lines. The perspective from Earth is shown in the top left panel, and two additional perspectives shown in the bottom two panels. A central point-light halo has been introduced to help distinguish front-facing from rear-facing structures. An enlargement of the jet is shown in the top right panel, which represents the region enclosed in the purple box in the top left panel. A transparent ellipsoid is added to represent the edge of the nebula body. The hole region (see section 4.3) is highlighted by the orange box in the top-right panel. An animation of Figure 3 is available (Movie 1). The first 28 seconds of the animation show one rotation around the north-south axis followed by one rotation around the east-west axis. The last 28 seconds show the same rotations but with the SN2 deep frame image superimposed on the model.}
\label{fig:oiii_whole}
\end{figure*}

The original 3D data cube for each emission line after the spectrum extractions contains over 18,000,000 individual data points from each filter. To enhance visualization, we performed the initial data cleaning step based on a flux signal-to-noise ratio (SNR) threshold of 10. 

Following the SNR cut, we applied a clustering algorithm to perform additional data cleaning. The clustering algorithm scans through each individual data point and identifies the four nearest neighbors. If all four nearest data points fall within a range of 1 arcsec (0.01 parsecs) from the scanned point, the data point is retained in the data set. Otherwise, the algorithm excludes the point and proceeds. 

The same SNR threshold and the clustering cleaning process were applied to the data from all filters, but some of the weak emission lines such as [\ion{N}{2}] $\lambda$5755 were selected based on an initial SNR threshold ranging from 5 to 10 before their clustering cleaning process.  

Data in the jet region required additional care. The [\ion{O}{3}] emission line flux in the northern jet region is much fainter compared to the nebula body, and the SNR threshold for the jet region was lowered to 2. Additionally, to achieve a more detailed data presentation at the breakout regions, the SNR threshold near the nebula's edge is set to 6, compared to the default threshold of 10 for the main nebula body. 

The final SNR cuts adopted and number of admissible data points after passing through our clustering algorithm for each filter are summarized in Table~\ref{cubes}. The data selection allows us to display regions of faint emission in the remnant on an equal footing with those of strong emission, providing a more complete view of the 3D visualization of the remnant. All cubes have the highest spectral and spatial resolutions of any previous survey of Crab Nebula at similar wavelength ranges \citep{Clark1983,Lawrence1995,Cadez2004,Martin2021}, and represent the most complete catalog of its optically emitting ejecta material to date.

\begin{figure*}
\centering
\begin{subfigure}{.45\textwidth}
  \centering
  \includegraphics[width=1\linewidth]{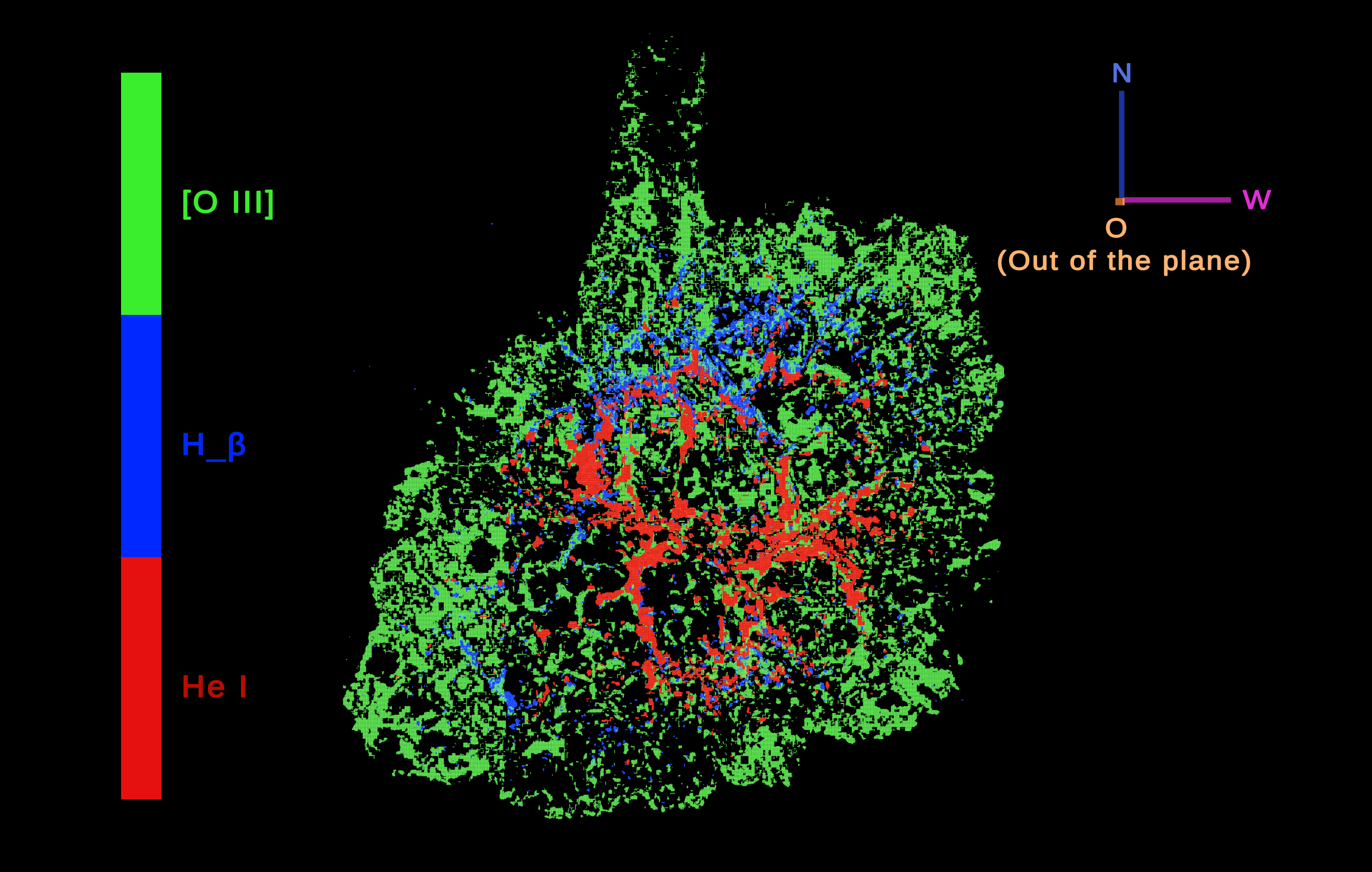} 
\end{subfigure}
\begin{subfigure}{.45\textwidth}
  \centering
  \includegraphics[width=1\linewidth]{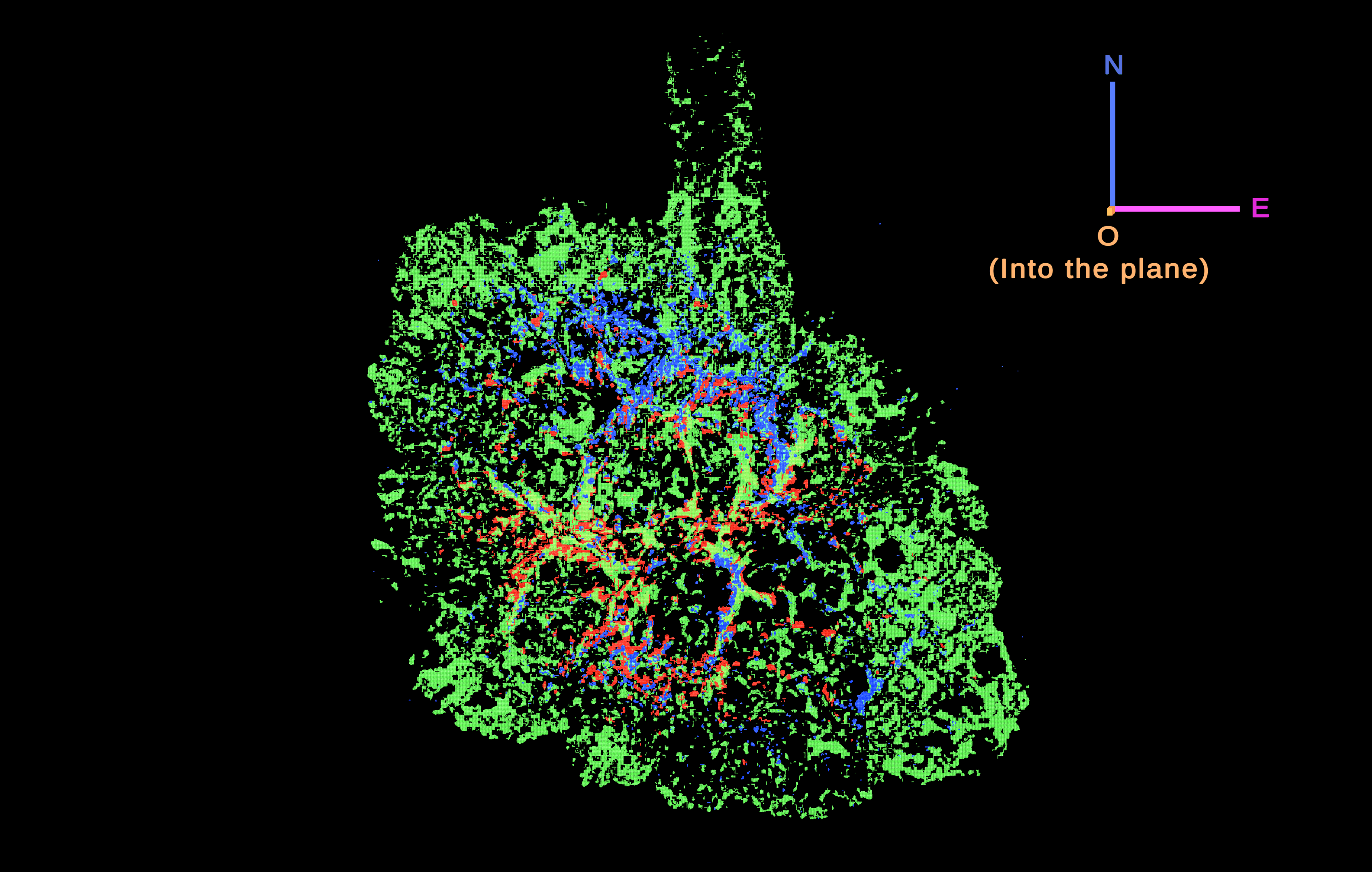}  
\end{subfigure}
\begin{subfigure}{.45\textwidth}
  \centering
  \includegraphics[width=1\linewidth]{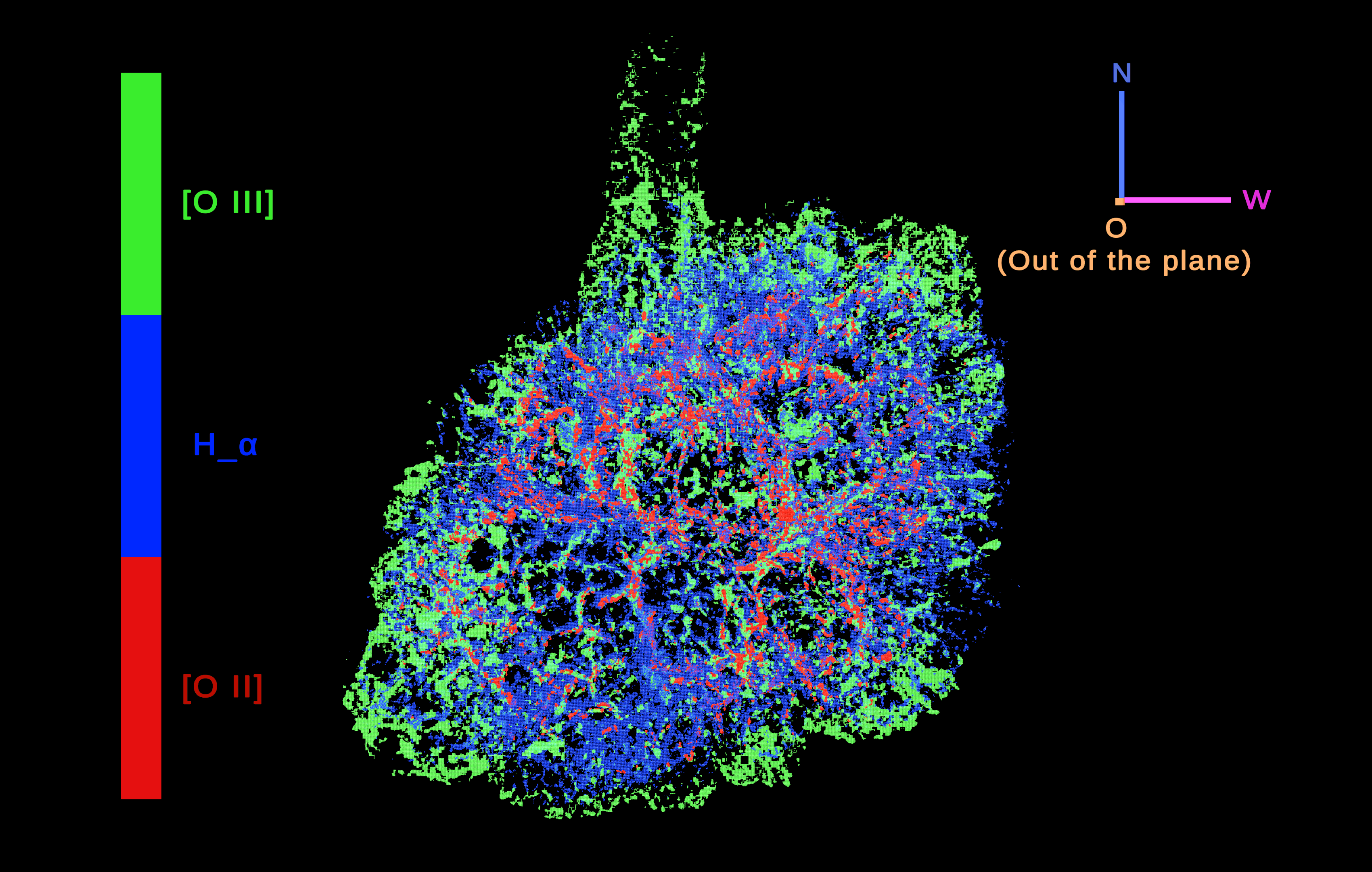}  
\end{subfigure}
\begin{subfigure}{.45\textwidth}
  \centering
  \includegraphics[width=1\linewidth]{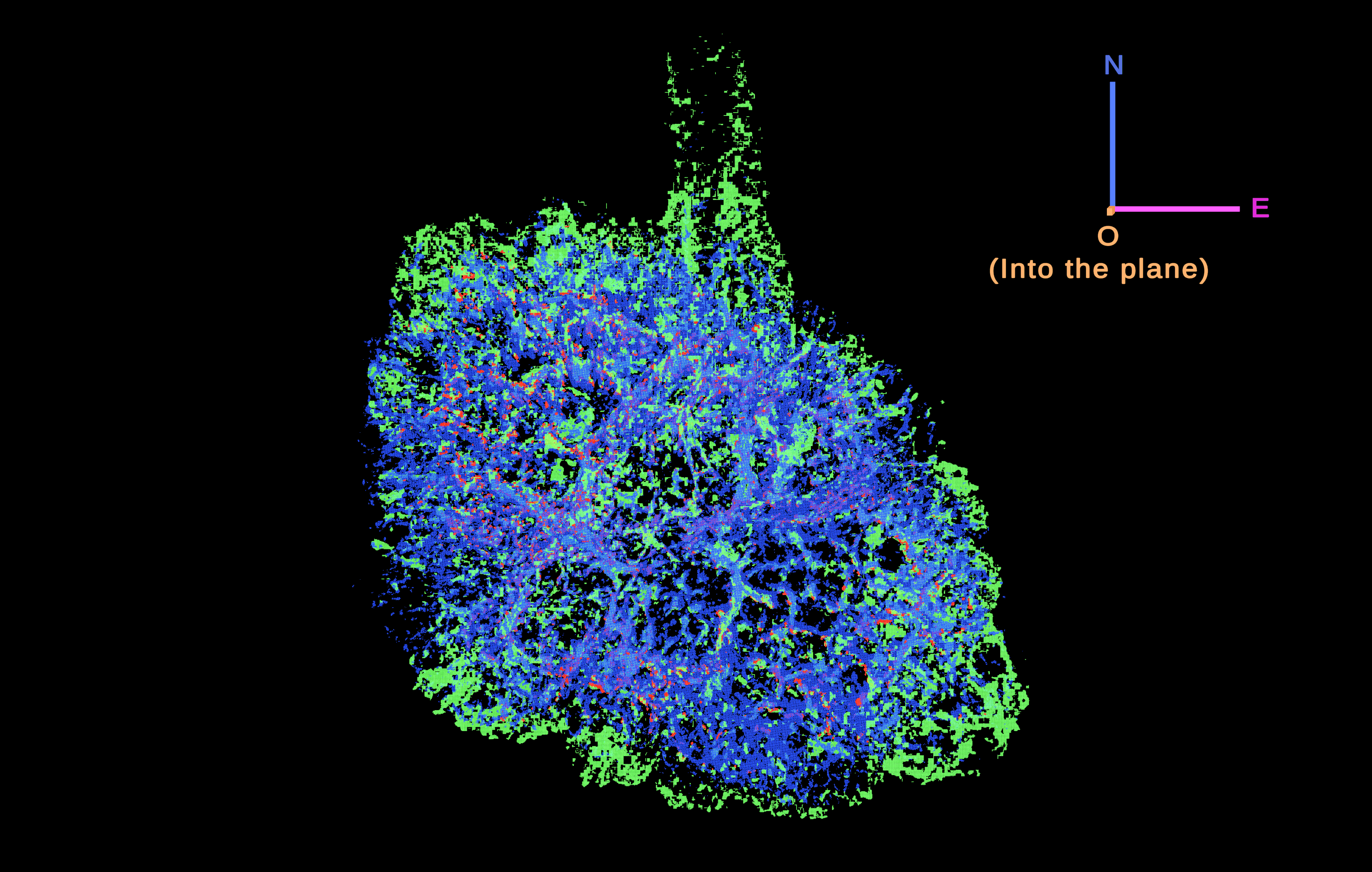}  
\end{subfigure}
\begin{subfigure}{.45\textwidth}
  \centering
  \includegraphics[width=1\linewidth]{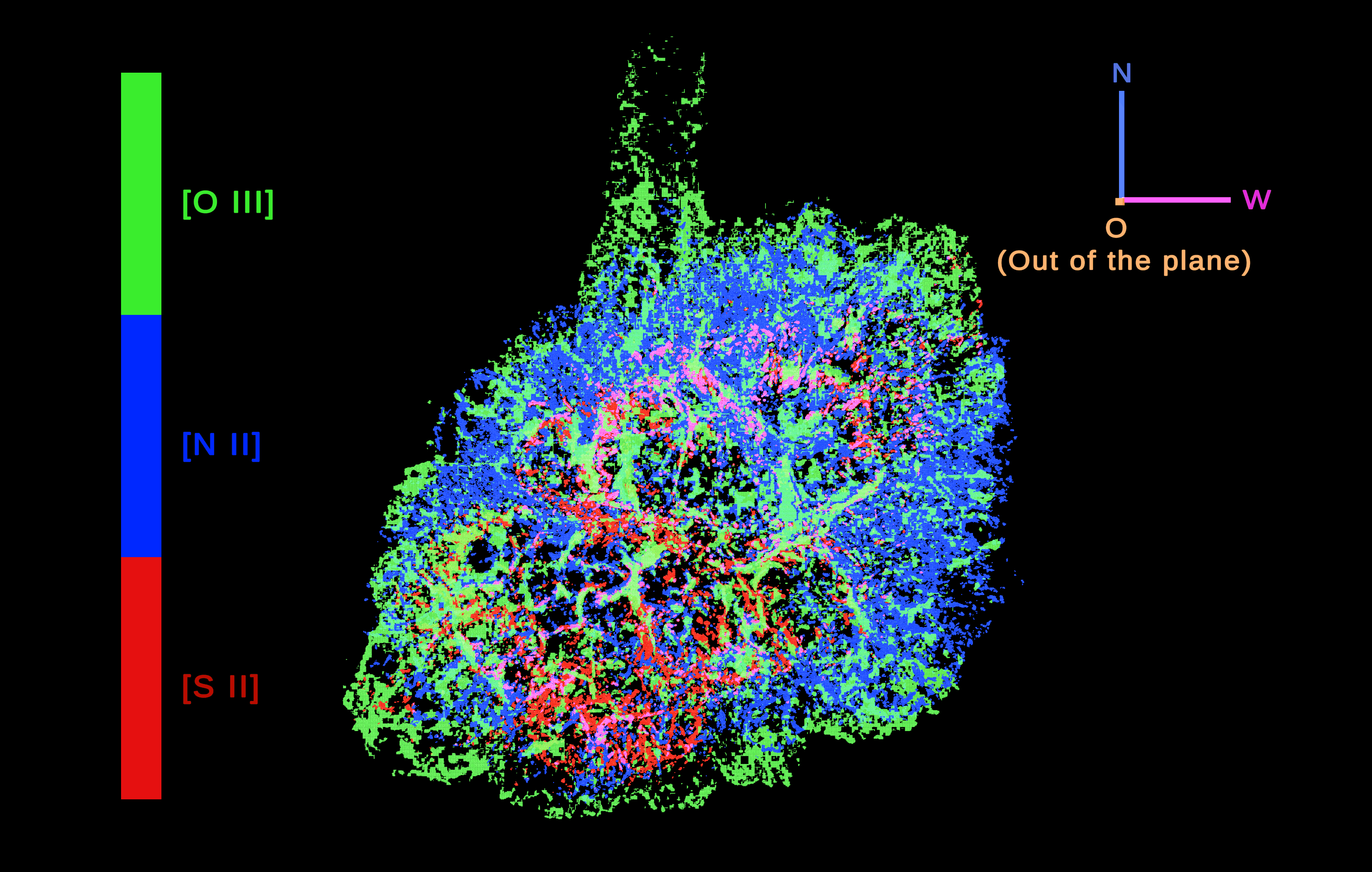}  
\end{subfigure}
\begin{subfigure}{.45\textwidth}
  \centering
  \includegraphics[width=1\linewidth]{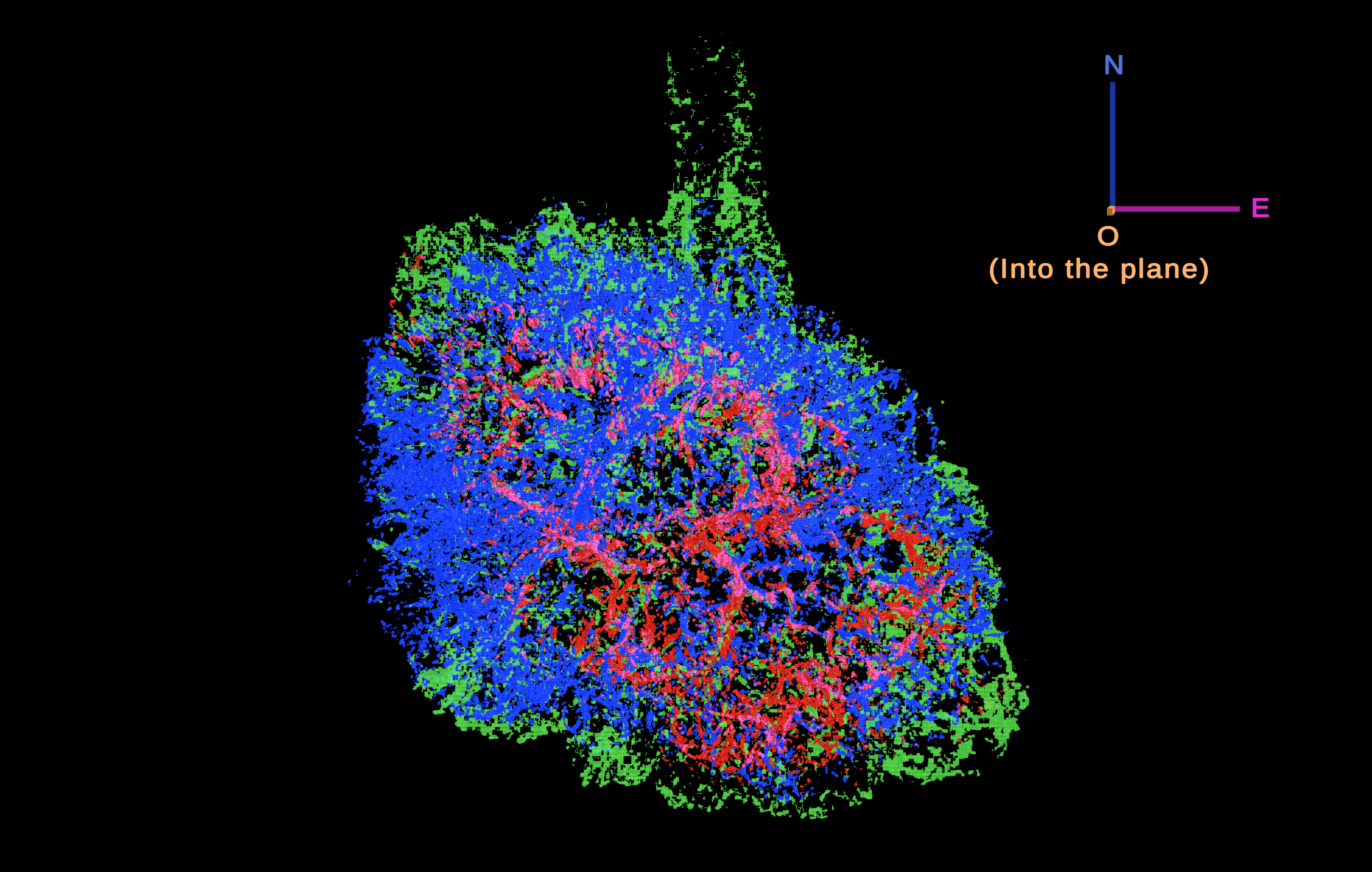}  
\end{subfigure}
\captionsetup{font=small}
\caption{3-dimensional reconstruction of Crab Nebula in [\ion{O}{2}] $\lambda$$\lambda$3726, 3729, [\ion{O}{3}] $\lambda$$\lambda$4959, 5007, $H\beta$, \ion{He}{1} $\lambda$5876, [\ion{N}{2}] $\lambda\lambda$6548, 6584, [\ion{S}{2}] $\lambda\lambda$6717, 6731, and $H\alpha$ emission lines. The Earth perspective is shown in the left panels. The right panels show the Crab rotated 180 degrees about the N-S axis. The majority of helium emission is concentrated inside the nebula body. An animation of Figure 4 is available (Movie 2). Every five seconds, the remnant rotates around its north-south axis while displaying a different emission line. During the final five seconds of the animation, the remnant is shown with all emission line components combined as it continues to rotate.}
\label{fig:chemical}
\end{figure*}

\section{Results} \label{sec:Results}

\subsection{3-D Reconstruction of the Crab Nebula} 

Figure~\ref{fig:oiii_whole} presents our 3-D reconstruction of the full Crab Nebula derived from the SN2 filter using [\ion{O}{3}] $\lambda$$\lambda$4959, 5007 emission. More than 360,000 individual data points are included, each color-coded by flux to highlight variations in surface brightness across the structure. To assist with depth perception, a central point-light halo is introduced to help distinguish front-facing from rear-facing structures without altering the underlying spatial distribution. An animation has also been created showing the individual data points rotated along the north-south and east-west axes, with and without the SN2 deep frame image (Movie 1). 

In H$\alpha$ emission, our new reconstruction confirms that the remnant exhibits the ``heart-shaped" morphology identified by \cite{Martin2021}. However, the structures revealed in other ions show significant departures from this geometry. These differences demonstrate that the Crab Nebula's three-dimensional structure varies markedly among emission species, reflecting the complex spatial distribution and excitation conditions of filaments traced by distinct ionic lines.  

The top right panel of Figure~\ref{fig:oiii_whole} enlarges the reconstruction around the jet region. Since part of the jet structure and a newly discovered cavity structure (later referred to as a ``hole" in the paper, see section \ref{subsec:Hole Region}) are challenging to see within the bulk of the ejecta, a fiducial ellipsoid was plotted in the figure to indicate their locations inside the nebula body. The hole structure is highlighted by the orange box. The enlarged view best illustrates that the orientation of the hole structure at the bottom of the jet and the titled angle of the jet funnel are well aligned. 

In Figure~\ref{fig:chemical}, we show three-dimensional reconstructions of several ionic species overlaid on the [\ion{O}{3}] $\lambda$$\lambda$4959, 5007 (green) distribution, including [\ion{O}{2}] $\lambda$$\lambda$3726, 3729 (red), \ion{He}{1} $\lambda$5876 (red), H$\beta$ (blue), H$\alpha$ (blue), [\ion{N}{2}] $\lambda\lambda$6548, 6584 (red), and [\ion{S}{2}] $\lambda\lambda$6717, 6731 (red). An animation has been made to show these data rotated individually and together along the north-south axis (Movie 2). 

Compared with the other emission lines, the \ion{He}{1} emission is centrally concentrated, occupying the innermost regions of the nebula where slower-moving material resides, while the oxygen and hydrogen lines extend throughout the full volume traced by the filamentary shell. This spatial concentration of helium is consistent with previous spectroscopic and photometric studies, which find that the Crab contains less than $\sim 2$\,M$_{\odot}$ of helium and that this component is associated with some of the slowest, most centrally confined ejecta \citep{MacAlpine1989, MacAlpine1991, MacAlpine2007}.

Interestingly, the Crab Nebula appears conspicuously compressed along the radial-velocity (LOS) dimension compared to its extent in the plane of the sky (see bottom right panel of Figure~\ref{fig:oiii_whole}). Assuming a distance of 2 kpc, we convert angular separations into physical positions in the plane of the sky. Adopting a smaller distance of 1.6 kpc, near the lower bound of recent Crab pulsar distance estimates \citep{Lin2023,CruzCruz2025}, would reduce the inferred plane of sky extent by approximately 20\%, bringing it into closer agreement with the nebula's projected size along the LOS direction.

Finally, we note that our three-dimensional reconstruction of the H$\alpha$ emission cube using the 2022 SN3 filter dataset represents a substantial improvement over the earlier 2016 reconstruction presented by \citet{Martin2021}; a direct comparison is provided in the Appendix.

\subsection{Jet Structure} \label{subsec:Jet Region}

\begin{figure}
    \centering
    \includegraphics[width=1\linewidth]{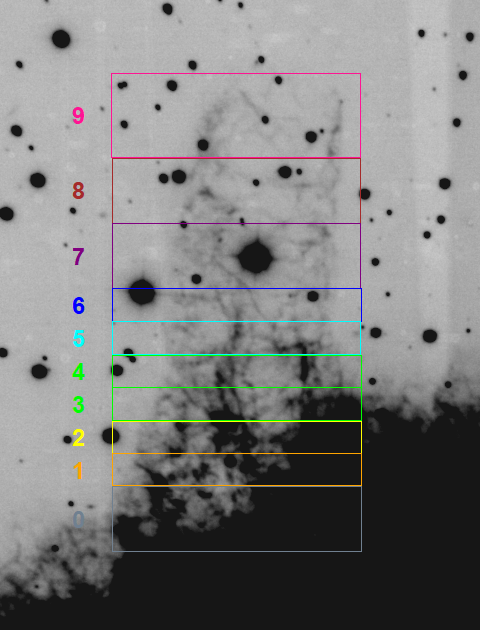}
    \caption{Positions of the jet layer segments. The bottom of layer 0 is 1.2 parsecs (120 arcsec) away from the nebula expansion center. Each layer covers 0.1 to 0.2 parsecs (10 to 20 arcsec) in height with the same 0.75 parsecs (75 arcsec) width. The top of layer 9 is 2.75 parsecs (275 arcsec) away from the nebula expansion center. The background image is an enlarged version of the SN2 filter deep frame shown in Figure \ref{fig:deepframe}.}
    \label{fig:jetds}
\end{figure}

To examine the structure of the jet funnel in detail, we divided it into ten sequential layers (Figure~\ref{fig:jetds}). Each layer spans 0.1 - 0.2 pc (10 - 20 arcsec) in height and shares a constant width of 0.75 pc (75 arcsec). The layers are numbered from 0 at the base of the jet to 9 at its northern tip. Layer 0 lies at the interface between the extended jet funnel and the main nebula body, positioned 1.2 pc (120 arcsec) from the nebula's center of expansion. The uppermost layer, Layer 9, reaches 2.75 pc (275 arcsec) from the expansion center, marking the top of the jet.

\begin{figure*}[ht]
    \centering
    \includegraphics[width=0.9\linewidth]{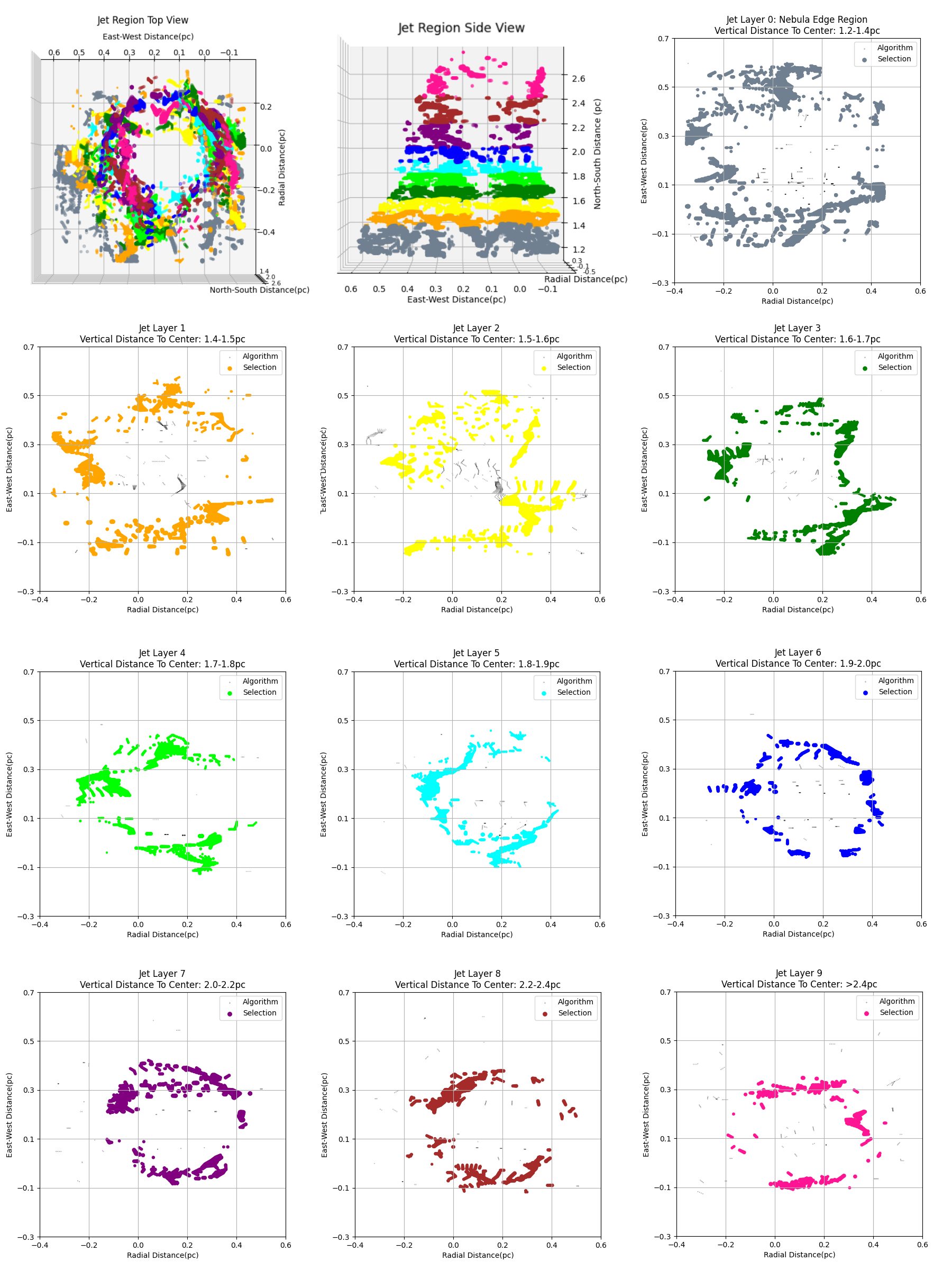}
    \captionsetup{font=small, skip=5pt}
    \caption{Layers of the jet funnel of the Crab Nebula. Numbering of layers increases with distance away from the main ejecta cage.  Selection of layers closely follows the approach used in Figure 4 of \cite{Black2015}. Data points not part of the jet funnel boundary are plotted as black dots. The size of each data point represents its flux relative to the highest flux value in the entire jet region, with larger points corresponding to higher flux values.}
    \label{fig:jetlayers}
\end{figure*}

For each jet layer, the clustering algorithm was performed first to remove noise and spurious detections. To outline the circular wall of the jet, we manually selected data points and applied color coding that emphasizes the ring structure of each jet layer (Figure~\ref{fig:jetlayers}). This selection closely follows the approach used by \cite{Black2015} to illustrate the funnel shape of the jet. All data points not selected for color coding are plotted as black dots in each layer. The size of each data point represents its flux relative to the highest flux value in the entire jet region, with larger points corresponding to higher flux values.

Moving northward along the jet, the emission strength steadily decreases. Starting from the fourth layer, 1.7 parsecs away from the expansion center, the elliptical shape of the jet wall becomes clear, which is consistent with the results of \cite{Black2015}. This elliptical ring is approximately 0.55 parsecs (55 arcsec or 600 km s$^{-1}$) length in the radial direction and a width of 0.4 - 0.5 parsecs (45 arcsec or 500 km s$^{-1}$) in the east-west direction. Most of the walls in the jet layers show a thickness of around 0.05 to 0.1 parsecs (5 - 10 arcsec or 50 - 100 km s$^{-1}$), which agrees with the results from \cite{Black2015}. We note that adopting a lower-limit distance of 1.6~kpc, as considered earlier, would cause these jet walls to appear even more elliptical in our 3D reconstruction than in the current version.

Although the adjacent layers show similar appearance and dimension, the jet walls are less complete and dimmer in some of the layers. The centers of the elliptical rings drift gradually to the west side of the nebula from the Earth's perspective and to the redshifted side of the radial direction. 

By comparing the center of the color-coded outer ring of each layer, we estimate the angle of inclination of the jet into the plane of the sky (red-shifted) to be roughly 8 $\pm$ 2 degrees, which is consistent with the result from \cite{shull1984} and 
\cite{Black2015}. Using the same center difference of each layer's outer ring, we also estimated the jet angle of inclination toward the west side of the nebula from the Earth's perspective to be 5 $\pm$ 2 degrees.

The first two plots in Figure~\ref{fig:jetlayers} show 3-dimensional perspectives of the entire jet data. Only the color-coded selection data for each layer were plotted in these figures to demonstrate the overall funnel shape of the jet.

Our results confirm that the bright emission in the Crab's northern jet forms a well-defined funnel-shaped structure, with the brightest filaments concentrated along its thin outer walls \citep{gull1982,veron1985,marcelin1990,rudie2008,Black2015}. Earlier investigations reported negligible line emission within the volume enclosed by these walls \citep{Black2015}, suggesting that the interior of the funnel contains little or no line-emitting material. Our survey enables a quantitative assessment of the line emission within this region, allowing us to directly test whether the interior of the funnel is indeed empty.

\begin{figure}[tp]
\includegraphics[width=\linewidth]{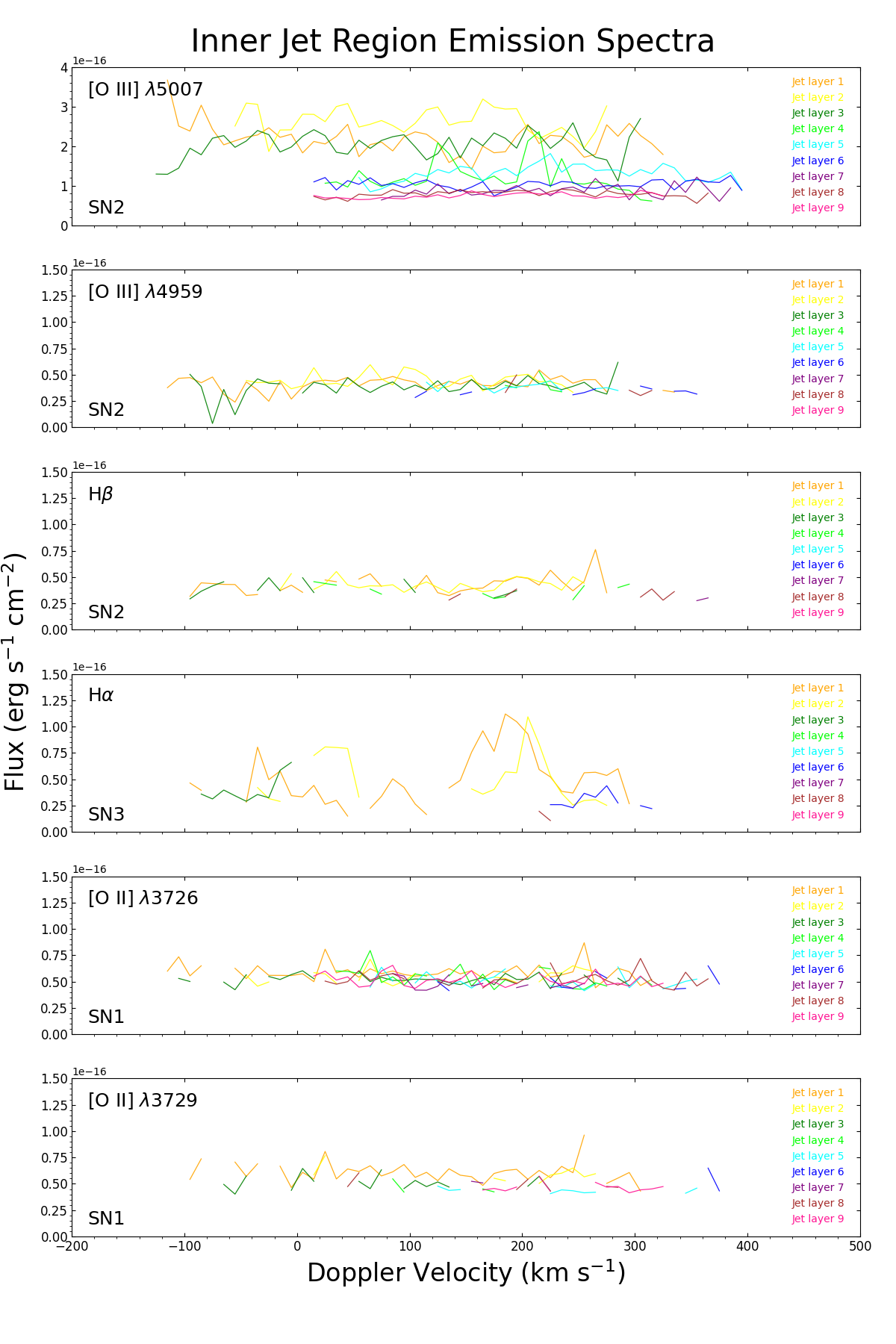}
\captionsetup{font=small}
\caption{Spectra of the inner jet-layer walls ejecta shown in different emission lines. The same colors are chosen to represent each jet layer matching Figure~\ref{fig:jetds} and Figure~\ref{fig:jetlayers}.}
\label{fig:jet_inner}
\end{figure}

Figure~\ref{fig:jet_inner} presents spectra of ejecta located within the jet funnel, spatially integrated across each reconstructed layer shown in Figure~\ref{fig:jetds}. All emission within the jet-layer walls, i.e., any emission located interior to the elliptical boundaries defined by ejecta labeled ``Selection", excluding regions around stars, is used to construct the integrated spectrum. The color scheme used for each inner jet-layer emission spectrum in Figure~\ref{fig:jet_inner} corresponds to that used for the respective jet-layer walls in Figures~\ref{fig:jetds} and~\ref{fig:jetlayers}, enabling a direct comparison between the spatial and spectral properties of material interior to and along the wall of the jet funnel.

\begin{figure}[tp]
    \centering
    \includegraphics[width=\linewidth]{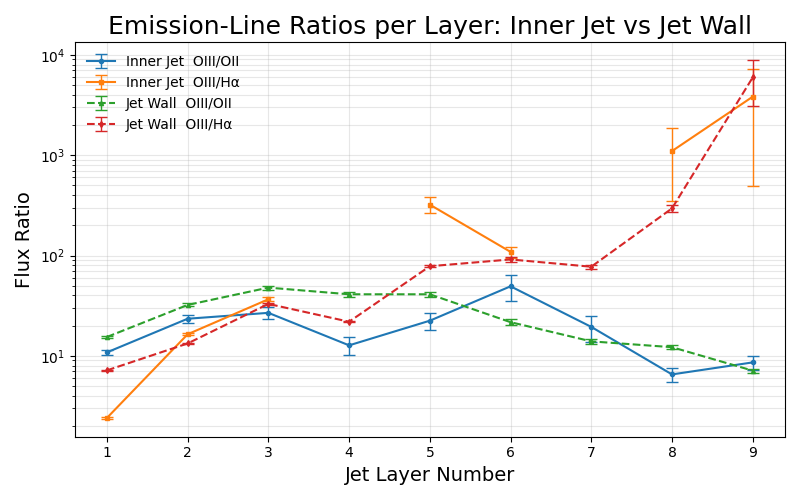}
    \caption{Flux ratios for emission lines in the jet as a function of layer away from the nebula, separated into two categories: the thin walls that define the jet (jet wall), and the volume interior to these walls (inner jet).}
    \label{fig:Flux_jetlayer}
\end{figure}

\begin{figure*}[tp]
\centering
\includegraphics[width=0.9\textwidth]{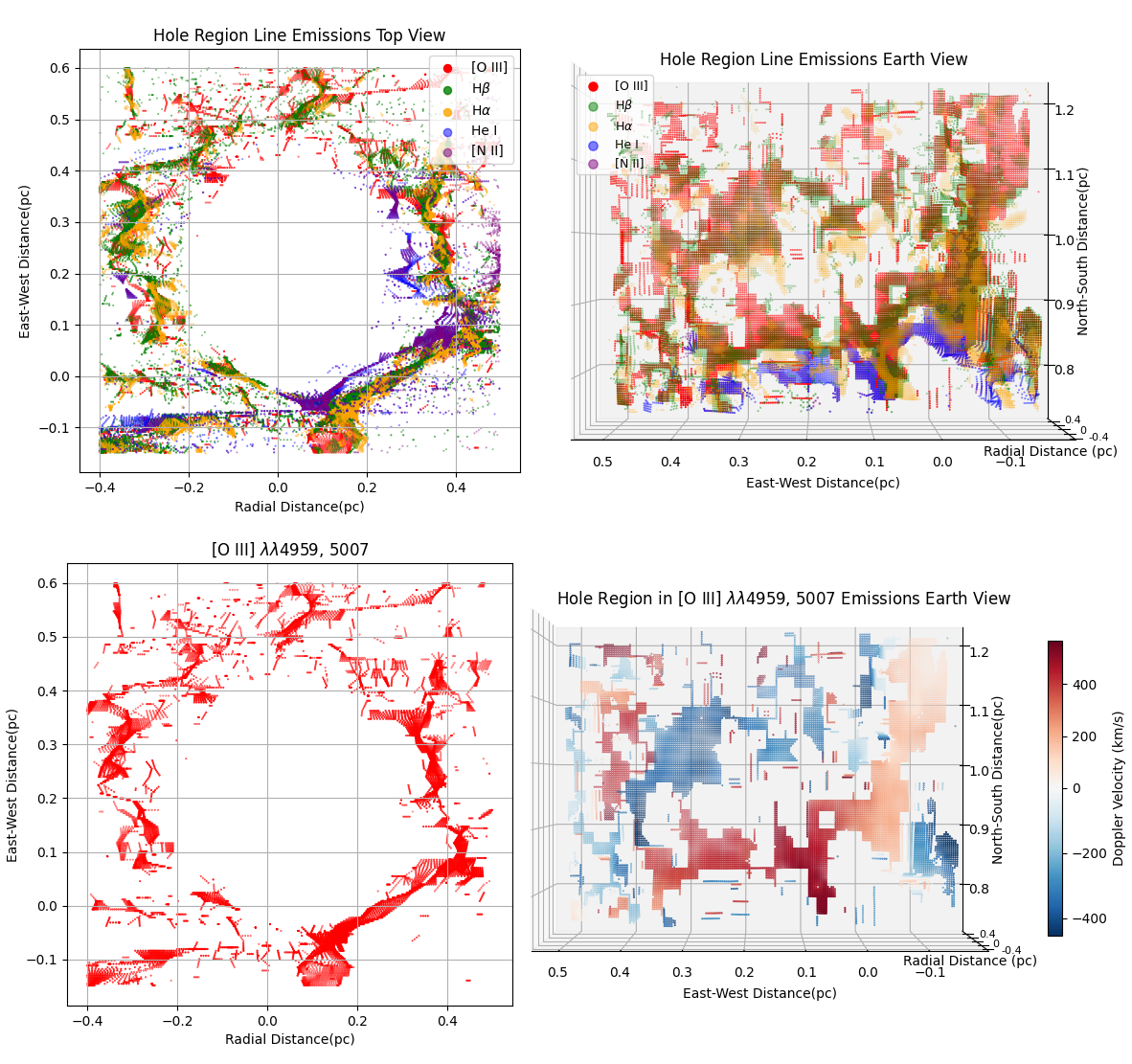}
\captionsetup{font=small, skip=4pt}
\caption{The cavity/hole region identified in the Crab Nebula at the base of the jet, shown in different emission lines. The same data processing algorithm has been used for each emission line. The size of each data point represents its flux relative to the highest flux value from the same emission line in the entire hole region, with larger points corresponding to higher flux values.}
\label{fig:hole_all}
\end{figure*}

\begin{figure*}[t]
    \centering
    \vspace{-1em} 

    \begin{subfigure}{\textwidth}
        \centering
        \includegraphics[width=0.9\linewidth,height=0.4\textheight,keepaspectratio]{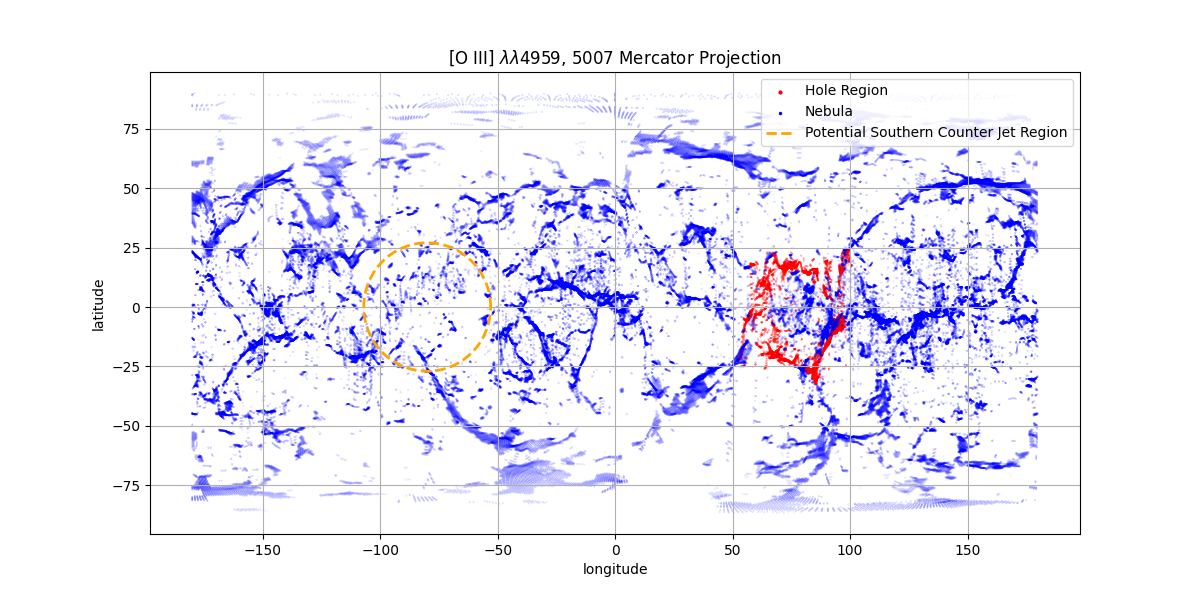}
    \end{subfigure}

    \vspace{-2.2em} 

    \begin{subfigure}{.9\textwidth}
     \centering
        \includegraphics[width=0.45\linewidth]{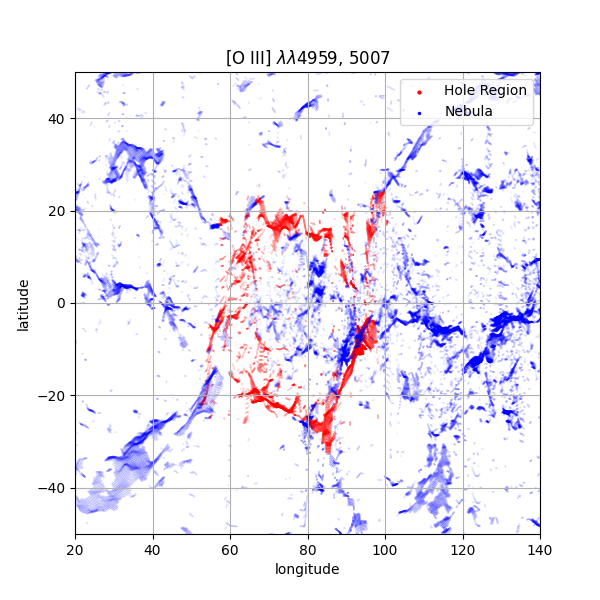}
        \includegraphics[width=0.45\linewidth]{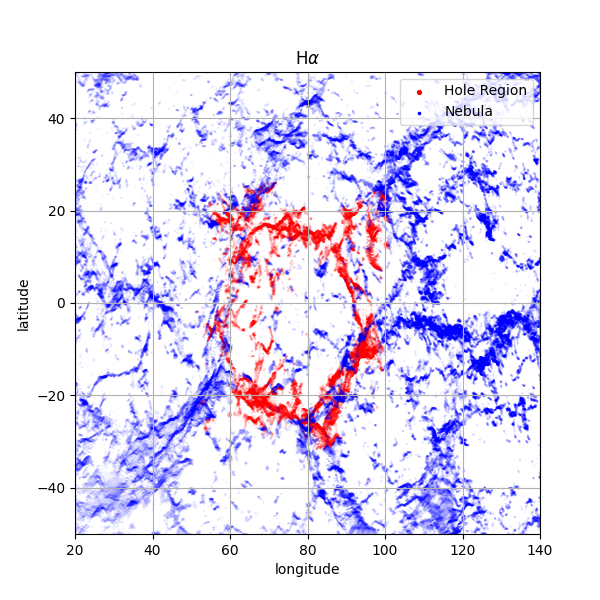}
    \end{subfigure}%

    \vspace{0.5em}

    \captionsetup{font=small, skip=5pt}
    \caption{Mercator projection of our 3-D reconstruction of the Crab Nebula. The hole region is highlighted in red (refer to Fig.~\ref{fig:deepframe}). An orange dashed circle is plotted to mark the region 180$^{\circ}$ opposite the northern jet, corresponding to a potential location if a southern counter-jet to the Crab's northern jet exists. Enlargements are shown in the bottom panels for both [\ion{O}{3}] and H$\alpha$, demonstrating how the walls of the cavity are observed across emission lines.}
    \label{fig:mercator}
\end{figure*}

Figure~\ref{fig:jet_inner} shows spectra of inner jet emission in velocity space for different ions: $H\beta$ and [\ion{O}{3}] $\lambda$$\lambda$4959, 5007 from SN2 filter, [\ion{O}{2}] $\lambda$$\lambda$3726, 3729 from SN1 filter, and $H\alpha$ from SN3 filter. For the C2 filter, emission is too weak to generate an integrated spectrum for material inside the jet funnel.

Most of the inner jet emissions are shown to be located near the base of the jet (Layer 1, 2, and 3). There are little to no $H\alpha$ emissions beyond the fourth layer of the jet. The oxygen emission flux in each layer of the inner jet region beyond layer 4 are approximately the same. 

In Figure~\ref{fig:Flux_jetlayer}, we compare the emission line flux ratios between the inner jet region and the jet wall region. There is no $H\alpha$ emission in the inner jet region for layer 4 and 7. Both the [\ion{O}{3}]/[\ion{O}{2}] and [\ion{O}{3}]/$H\alpha$ ratios show the same trend for all nine layers when comparing the inner jet with the jet wall region, suggesting that there are no significant differences between the two regions. The integrated flux of each layer in the jet wall region is $\sim10^{-12}$ erg s$^{-1}$, roughly ten times higher than in the inner jet layers, indicating that most of the northern jet's total emission is concentrated in the jet wall region.

\begin{figure*}
\begin{subfigure}{.5\textwidth}
  \centering
  \includegraphics[width=1\linewidth]{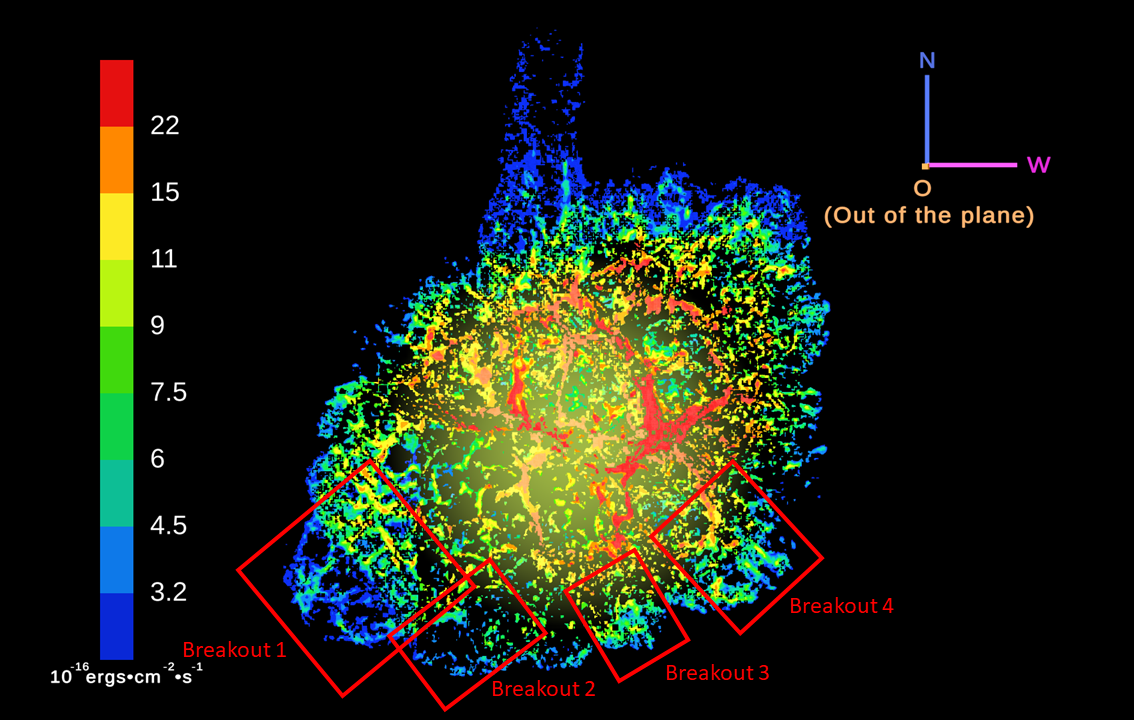}  
\end{subfigure}
\begin{subfigure}{.5\textwidth}
  \centering
  \includegraphics[width=1\linewidth]{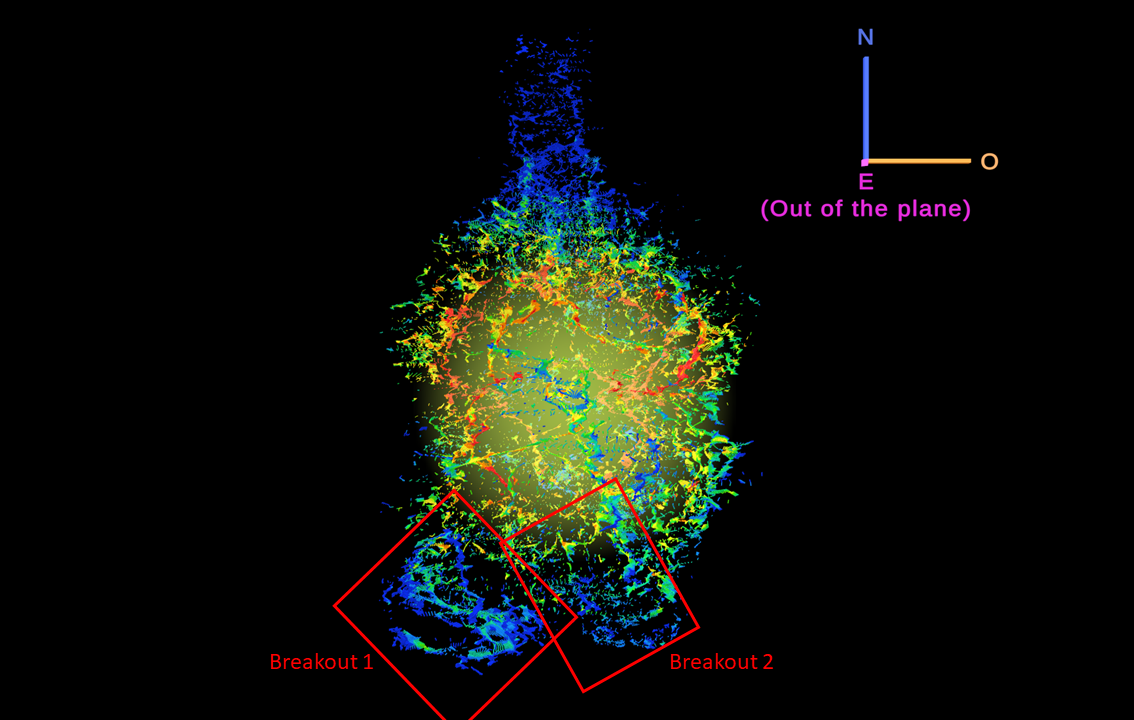}  
\end{subfigure}
\begin{subfigure}{.5\textwidth}
  \centering
  \includegraphics[width=1\linewidth]{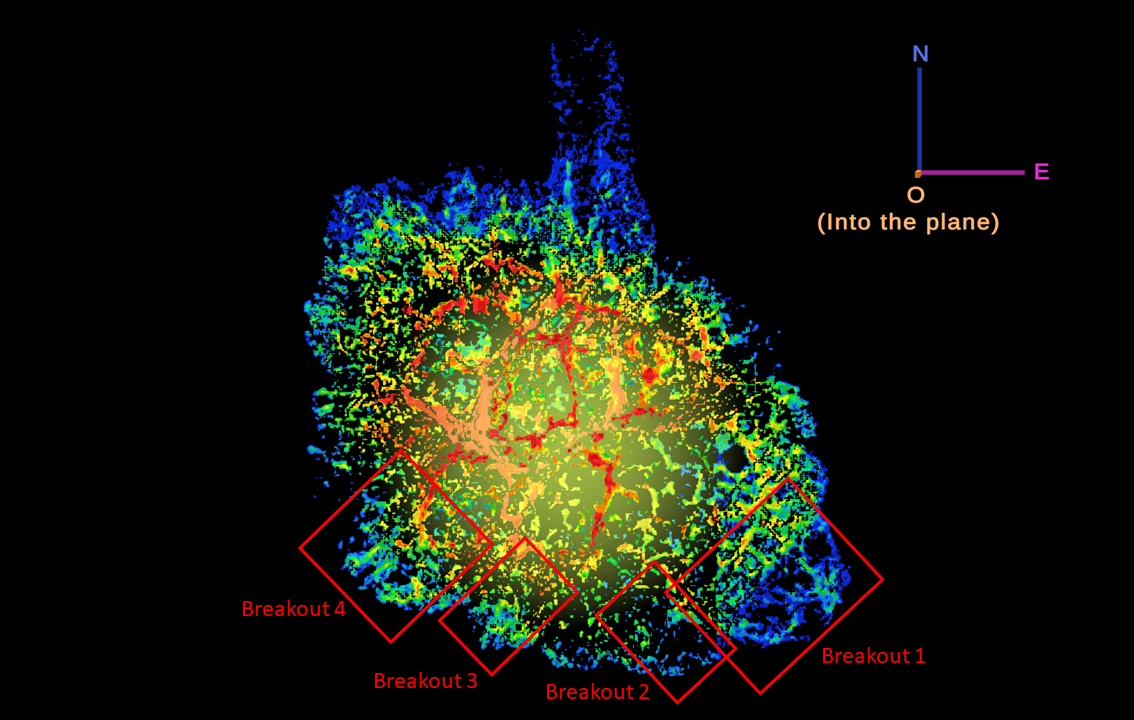}  
\end{subfigure}
\begin{subfigure}{.5\textwidth}
  \centering
  \includegraphics[width=1\linewidth]{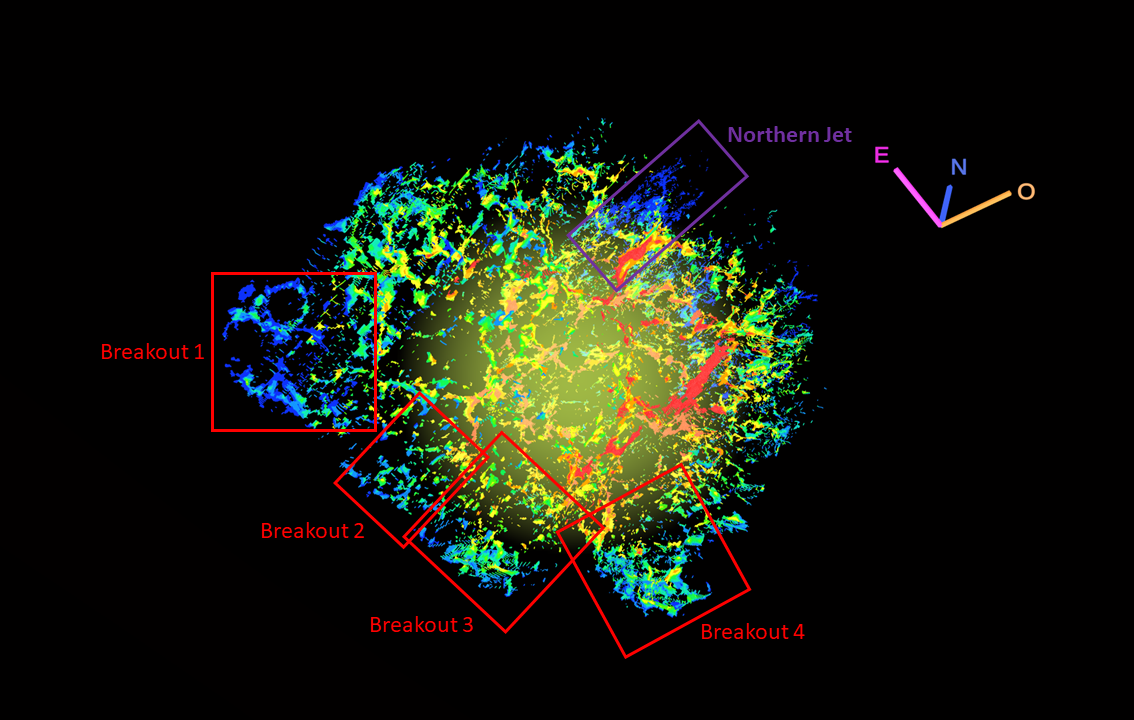} 
\end{subfigure}
\captionsetup{font=small}
\caption{Four of the most conspicuous breakouts seen in our 3-D reconstruction of Crab Nebula in [\ion{O}{3}] $\lambda$$\lambda$4959, 5007 emission lines where filaments extend out noticeably from the main cage of emission.}
\label{fig:breakouts}
\end{figure*}

\subsection{Hole/Cavity in the Nebula\label{subsec:Hole Region}}

\cite{Black2015} identified a long and fairly distinct area nearly devoid of emission under the jet region (see their Figure 8). This area, called the ``channel,'' was discovered between the jet and the remnant's expansion center (see also \citealt{fesen1997}). 

Our complete mapping of the entire Crab Nebula and jet in [\ion{O}{3}] shows this unambiguously, for the first time in 3-D, as a connecting region between the ejecta filaments and jet.  The detail of our 3-dimensional reconstruction shows that this channel is actually an annular structure, which we refer to as the ``hole.'' The hole is part of the emission shell of the Crab, directly beneath the jet region, and is approximately 0.9 parsec (90 arcsec) away from the nebula's expansion center. To enhance the hole representation, we followed a similar procedure as that used for the jet: an initial SNR cut was applied before implementing the clustering algorithm, then a manual selection was made to highlight the annular structure. After cleaning, the annular structure within the hole region exhibits a nearly circular shape, characterized by a radius ranging from 0.3 to 0.4 parsecs (30 to 40 arcsec or 300 to 400 km s$^{-1}$), with a thin wall that has a width of approximately 0.03 parsecs (3 arcsec or 30 km s$^{-1}$). The integrated flux of the emission from the hole wall is $\sim 9\times10^{-13}$ erg s$^{-1}$ for each emission line, approximately three times higher than that of the inner hole region. Additional details about our processing of the hole can be found in the Appendix.

The annular structure seen in [\ion{O}{3}] $\lambda$$\lambda$4959, 5007, is observed in many emission lines, particularly H$\alpha$ and H$\beta$. Figure~\ref{fig:hole_all} shows the hole region in multiple emission lines after the same data reduction processes. The size of each data point represents its flux relative to the highest flux value from the same emission line in the entire hole region, with larger points corresponding to higher flux values. The fact that the hole can be seen in different line emissions supports the conclusion that this is a true coherent structure of SN ejecta.

The top right panel of Figure~\ref{fig:hole_all} displays a side view from the Earth's perspective of the hole region with all emission lines included. Shown in the bottom right panel, the blue-side semicircle of the hole lies around 0.2 parsec (20 arcsec) higher than the red-side semicircle, resulting in an estimated tilt of about 14 degrees into the plane of the sky. The west-side semicircle of the hole is farther from the expansion center than the east-side semicircle. This orientation aligns with the direction of the jet funnel discussed above.

We provide an additional perspective of the hole region in Figure~\ref{fig:mercator}, which shows Mercator projections of the entire Crab. We adopt the expansion center as the origin of the Mercator projection. In this coordinate system, longitude corresponds to the azimuthal angle in the plane of the sky, where $0^\circ$ points toward the east side of the remnant, $90^\circ$ toward the north, $-90^\circ$ toward the south, and $\pm180^\circ$ toward the west. The latitude traces the LOS direction, distinguishing between red-shifted and blue-shifted material. In this convention, the north-pole region corresponds to the front side of the remnant, while the south-pole region corresponds to the back side. An orange dashed circle is plotted to mark the region 180$^{\circ}$ opposite of the northern jet, corresponding to a potential location if a southern counter-jet to the Crab's northern jet exists. Also presented are enlargements of the Mercator projection around the hole to highlight its coherent structure. Several ring-like structures of various sizes are observed in other locations of the nebula. None, however, share the defining characteristics of the hole region: coherent structure extending radially outward over a length comparable to the radius of the remnant itself.

\subsection{Filament ``Breakouts"}

In addition to the northern jet, several regions around the Crab Nebula show filamentary structures extending noticeably beyond the main ejecta cage of the remnant. These protrusions, or ``breakouts,'' are most clearly visible in the [\ion{O}{3}] $\lambda$$\lambda$4959, 5007 emission. Two breakouts appear along the southeastern limb, another lies just west and slightly below the base of the northern jet, and a fourth is located along the southwestern edge of the nebula. Four different viewing angles of our 3-D reconstruction highlight the locations and extents of these breakout features in Figure~\ref{fig:breakouts}. 

Breakout 1 is strongly tilted into the plane of the sky and is the largest and most extended of the four features. Breakout 2 lies on the blueshifted side of the southern limb and is much fainter, with a less complete funnel structure, smaller radius, and shorter distance from the expansion center than Breakout 1. Breakout 3 is located slightly west and below the northern jet, with an orientation that also points modestly westward. Breakout 3 has a flux roughly twice that of Breakouts 1 and 2. Breakout 4 likewise exhibits relatively high flux and is directed toward the west-south side of the nebula. Its radius is nearly twice that of Breakout 3.

\begin{figure*}
\begin{subfigure}{.5\textwidth}
  \centering
  \includegraphics[width=1\linewidth]{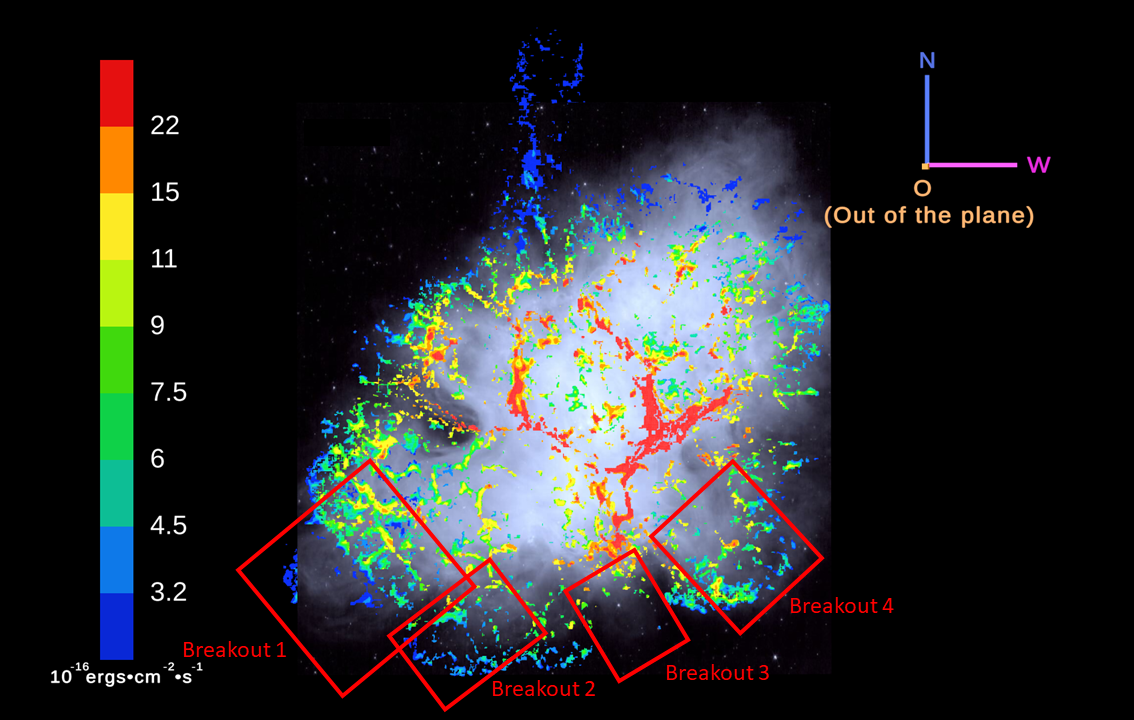}  
\end{subfigure}
\begin{subfigure}{.5\textwidth}
  \centering
  \includegraphics[width=1\linewidth]{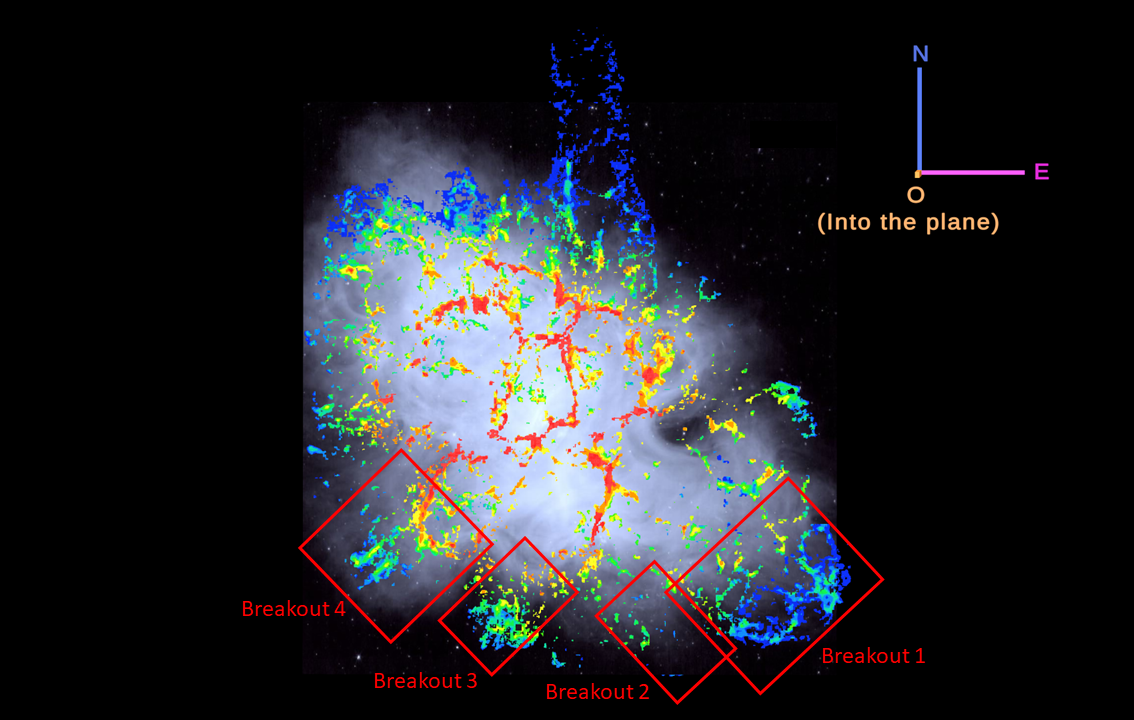}  
\end{subfigure}
\captionsetup{font=small}
\caption{Same [\ion{O}{3}] $\lambda$$\lambda$4959, 5007 3-dimensional reconstruction of Crab Nebula as Figure~\ref{fig:breakouts}, but with a JWST/MIRI image (F480M) from \citet{Temim2024} included to demonstrate how breakouts are correlated with extensions of the Crab synchrotron nebula. An animation of Figure 12 is available (Movie 3). The remnant rotates around its north-south axis, followed by a rotation around its east-west axis with the JWST image superimposed on the model. The animation pauses for a few seconds at several viewing angles to allow the viewer to better examine the breakout regions.}
\label{fig:jwst}
\end{figure*}

The breakouts appear to be correlated with synchrotron emission. In Figure~\ref{fig:jwst}, we plot our [\ion{O}{3}] emission 3D reconstruction with the new JWST observation of the Crab synchrotron nebula obtained with NIRCam using the F480M filter \citep{Temim2024}. An animation showing the [\ion{O}{3}] emission 3D reconstruction and these breakouts rotated about the north-south and east-west axes with respect to the JWST observation has been provided (Movie 3).

Breakout 4 correlates with an extension of the synchrotron nebula along the same direction as the pulsar disk. There is also enhanced synchrotron emission near the breakout 2 region. Breakout 1 aligns with a conspicuous extension of the synchrotron nebula, along the pulsar jet direction. Breakout 3 aligns with a small extension of the synchrotron nebula below the pulsar. 

In the breakout regions, the expanding PWN deposits energy into the surrounding ejecta, driving the [\ion{O}{3}]-emitting filaments beyond the remnant's boundary. Along the northwest edge, however, previous studies have shown that the PWN has already broken through this confining [\ion{O}{3}] shell \citep{Velusamy1984, hester2008crab}, and the synchrotron nebula has advanced ahead of the filamentary material. The absence of enhanced [\ion{O}{3}] emission along the northwest edge of the synchrotron nebula is therefore consistent with a region where the PWN has already escaped its ejecta cage rather than one where it is currently driving a breakout.

\section{Discussion}
\subsection{Jet and Hole}

\begin{figure*}[tp]
    \centering
    \includegraphics[width=0.8\linewidth]{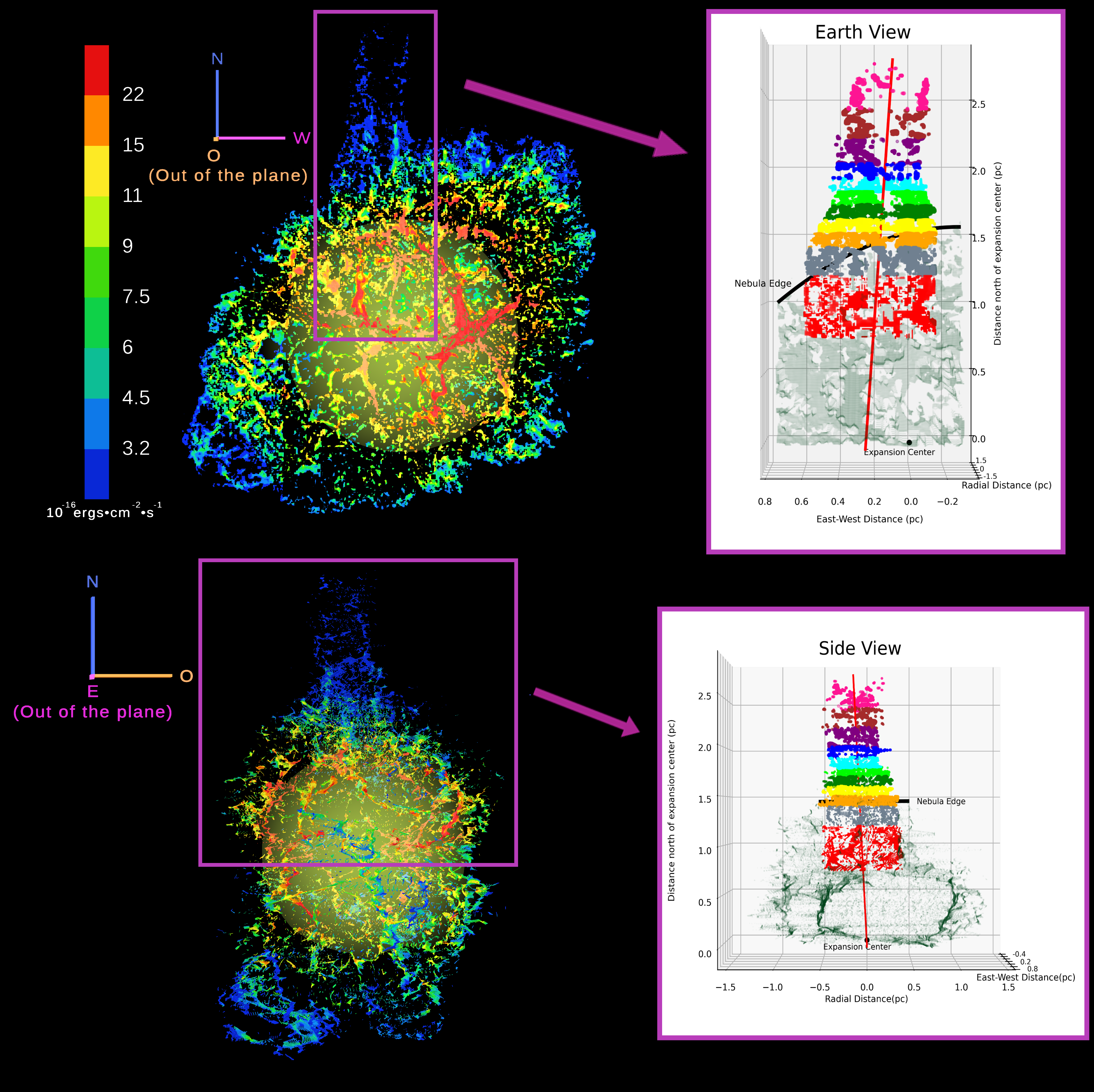}
    \captionsetup{font=small}
    \caption{Enlargement of the jet and hole region, emphasizing how the orientation of a cavity in the main cage of emission is aligned with the jet funnel. The colors of each jet layer and the hole/cavity region are the same as those used in Figure~\ref{fig:jetlayers} and Figure~\ref{fig:hole_all}. Other [\ion{O}{3}] emission data points near the jet and hole regions are plotted in dark green to illustrate how deeply the jet-cavity structure extends into the edge of the remnant. A red line connects the center of the top jet layer (pink) to the center of the hole/cavity (red), and extends toward the expansion center of the nebula (black dot). From the Earth's perspective, the red line lands 0.25 parsec (25 arcsec) offset to the east of the nebula's expansion center. However, from a side perspective (here shown as looking east from the west), the red line ends up precisely at the expansion center.}
    \label{fig:jetholeflux}
\end{figure*}

Many aspects of the jet and the nebular hole/cavity suggest that they share a common origin. The two structures have comparable sizes, indicating that they may have been shaped by the same physical process. The cavity is positioned directly beneath the lowest layer of the jet, creating a clear spatial connection between the two features. In addition, both the jet and the cavity are aligned in the same direction, which further supports the idea that they formed together as part of a single, continuous structural pathway.

In Figure~\ref{fig:jetholeflux}, we present enlarged three-dimensional views of the jet funnel and its placement relative to the cavity and the surrounding region. These vantage points reveal how deeply the combined jet-cavity structure extends into the main filamentary cage of the Crab Nebula, cutting through what is otherwise a densely woven network of ejecta. The cavity itself has a radius of roughly 0.3 pc (30$^{\prime\prime}$), and the base of the jet extends about 0.6 pc (60$^{\prime\prime}$), demonstrating that these two features are comparable in scale. The close spatial alignment between the cavity and the lowest portion of the jet, together with their similar dimensions, underscores their physical connection. This geometry suggests that the jet is not an isolated outflow but instead emerges along a pre-existing channel that opens directly into the cavity, helping establish the continuous, funnel-like structure seen in our reconstructions.

A key aspect of interpreting the jet-cavity system is understanding how its axis aligns with the nebula's center of expansion, because this geometric relationship directly informs its underlying formation mechanism. As shown in Figure~\ref{fig:jetholeflux}, a red guideline drawn through the center of the cavity and the centers of the uppermost jet layers provides a visual reference for this alignment. When viewed in the plane of the sky from Earth, this guideline axis does not pass directly through the expansion center (marked by the black dot). Instead, it is offset by roughly 0.25 pc (25$^{\prime\prime}$). However, when the reconstruction is rotated to a side-on perspective that emphasizes depth along the LOS, the jet and cavity come into alignment with the expansion center.  This confirms that the offset between the jet-cavity axis and the expansion center is confined to the plane of the sky, with no significant displacement along the LOS. The origin of this lateral misalignment may reflect asymmetries in the pulsar wind nebula or circumstellar medium, either of which could drive asymmetric expansion of the nebula and thereby shift the jet-cavity axis away from the apparent center of expansion.

\subsection{The Origin of Northern Jet and Hole}

The origin of the Crab northern jet has been a topic of long-standing debate and the cavity revealed for the first time in our 3-D reconstruction provides an important new way to evaluate competing explanations for the origin of the Crab's northern jet. Our analysis shows that the cavity and the jet are closely linked and likely share a common origin. 

Against this backdrop, we now revisit the main theoretical explanations proposed for the origin of the Crab's northern jet, which are summarized in Table~\ref{origins}. (1) A mass-loss trail created by the red-giant progenitor, which could have carved a low-density channel prior to the explosion \citep{blandford1983, Cox1991}. (2) Magnetically shaped filaments, in which expanding magnetic fields sculpt the jet out of the surrounding ejecta  \citep{bychkov1975,marcelin1990}. (3) PWN-driven breakout models, where instabilities in the pulsar wind nebula create a funnel that continues the [\ion{O}{3}] skin outward skin \citep{chevalier1975,sankrit1997,Smith2013,Blondin2017}. (4) Rayleigh-Taylor breakout into a low-density ISM, allowing ejecta to penetrate the shell more easily in certain directions \citep{kundt1983,veron1985,Porth2014}. (5) High-velocity ejecta from the Crab's N-S bipolar expansion, supported by the orientation of the synchrotron ``bays'' \citep{Fesen1992}, and helium-rich filament populations \citep{MacAlpine1989,Fesen1992,Lawrence1995,Satterfield2012}, as well as direct kinematic arguments \citep{fesen1993,fesen1997,rudie2008, Black2015}. (6) Interaction with a local interstellar cloud, which could have blocked part of the expanding shell and redirected material into a jet-like structure \citep{morrison1985}. (7) A relativistic pulsar plasma beam, in which the pulsar produces a directed outflow that shapes or powers the observed jet \citep{shull1984,benford1984,michel1985,bietenholz1990}.

In the discussion that follows, we evaluate these scenarios in light of several additional observational constraints. These include the absence of a southern counter-jet, the pronounced collimation of the northern jet, the near-ballistic expansion of the ejecta, proper-motion measurements indicating an early formation time of the northern jet, roughly coeval with the nebula \citep{rudie2008}, and the close relationship between the jet and the Crab's [\ion{O}{3}] skin, particularly when considered alongside the breakout regions identified earlier. Taken together, these factors provide a coherent framework for assessing which formation mechanisms are most consistent with the observed structure and kinematics of the Crab's northern jet.

Among all the scenarios considered, Model (1) produces an annular structure as the progenitor star travels through the ISM prior to the explosion. However, it remains unclear how close such a mass-loss trail can form to the star before the explosion, or whether the explosion would destroy such an annular structure created by the star trail. Model (2) does not require or predict the formation of a cavity embedded inside the remnant's [\ion{O}{3}] skin layer. Model (3) requires PWN breaking out from the low density explosion ejecta region to form the jet. It remains unclear whether ejecta accelerated by the PWN would leave behind an annular structure. Models (4) and (6) treat the jet as an extension of the [\ion{O}{3}] skin or as material outside the nebula boundary, whereas the cavity we identify is located beneath the [\ion{O}{3}] layer and therefore cannot be produced by these mechanisms. As for model (5), the geometry that would be associated with the fastest explosion ejecta is uncertain, and the annular cavity that we observe in our 3D reconstructions may be consistent with model predictions. Model (7) also produces a cone-shaped/limb-shaped structure with the pulsar beam propagating outward along a cylindrical geometry defined by the magnetic field \citep{benford1984}, which accounts for the presence of the internal annular cavity we observe in our 3D reconstructions.

Most models can, in principle, explain the absence of a southern counter-jet, but model (7) faces challenges. The pulsar beam model generally produces two jets, one from each polar cap \citep{benford1984}. Similarly, the magnetic fields generated by the pulsar generally exhibit a bipolar structure, but it may vary at different directions in the case of model (2). For model (5), the fastest ejecta expanding southward in the remnant will expand into the region where the pulsar jet along the south-east axis points. The strong pulsar wind along this axis can redirect the fastest ejecta, causing them to pivot and become more closely aligned with the pulsar jet axis along the NW-SE direction. 

\begin{deluxetable*}{lcccccc}
\centering
\caption{\label{origins}}

Summary of proposed models for the origin of the Crab Nebula's northern jet and their ability to account for key observed properties.

\vspace{0.5em}

\tablewidth{0pt}
\renewcommand{\arraystretch}{2.5} 

\tablehead{
\colhead{Theory} & 
\colhead{\parbox[c]{1.4cm}{\centering \textbf{Presence\\of the Hole}}} & 
\colhead{\parbox[c]{1.6cm}{\centering \textbf{Lack of\\Southern\\Counter-Jet}}} & 
\colhead{\parbox[c]{1.6cm}{\centering \textbf{Different from\\Other\\Breakouts}}} & 
\colhead{\parbox[c]{1.5cm}{\centering \textbf{Early\\Formation\\Time}}} & 
\colhead{\parbox[c]{1.7cm}{\centering \textbf{Jet\\Cylindrical\\Shape}}} & 
\colhead{\parbox[c]{1.5cm}{\centering \textbf{Ejecta\\Near-Ballistic\\Expansion}}} 
\\[-1.8em] 
}
\startdata
\parbox[c]{5.2cm}{\centering (1) Mass-loss Trail} & \textbf{?} & \textbf{\textcolor{GoodMark}{\ding{51}}} & \textbf{\textcolor{GoodMark}{\ding{51}}} & \textbf{\textcolor{GoodMark}{\ding{51}}} & \textbf{\textcolor{GoodMark}{\ding{51}}} & \textbf{?} \\ [0.8em]
\tableline
\parbox[c]{5.2cm}{\centering (2) Filaments in Expanding\\Magnetic Field} & \textbf{\textcolor{BadMark}{\ding{55}}} & \textbf{?} & \textbf{\textcolor{BadMark}{\ding{55}}} & \textbf{\textcolor{GoodMark}{\ding{51}}} & \textbf{\textcolor{GoodMark}{\ding{51}}} & \textbf{?} \\ [0.8em]
\tableline
\parbox[c]{5.2cm}{\centering (3) PWN Instability Breakout} & \textbf{?} & \textbf{\textcolor{GoodMark}{\ding{51}}} & \textbf{?} & \textbf{\textcolor{GoodMark}{\ding{51}}} & \textbf{?} & \textbf{?} \\ [0.8em]
\tableline
\parbox[c]{5.2cm}{\centering (4) Expansion into a Low Density\\ISM Region} & \textbf{\textcolor{BadMark}{\ding{55}}} & \textbf{\textcolor{GoodMark}{\ding{51}}} & \textbf{\textcolor{GoodMark}{\ding{51}}} & \textbf{\textcolor{BadMark}{\ding{55}}} & \textbf{\textcolor{BadMark}{\ding{55}}} & \textbf{\textcolor{BadMark}{\ding{55}}} \\ [0.8em]
\tableline
\parbox[c]{5.2cm}{\centering (5) Highest-Velocity Ejecta of\\N-S Bipolar Expansion} & \textbf{?} & \textbf{?} & \textbf{\textcolor{GoodMark}{\ding{51}}} & \textbf{\textcolor{GoodMark}{\ding{51}}} & \textbf{?} & \textbf{\textcolor{GoodMark}{\ding{51}}} \\ [0.8em]
\tableline
\parbox[c]{5.2cm}{\centering (6) Interaction with a Local\\Interstellar Cloud} & \textbf{\textcolor{BadMark}{\ding{55}}} & \textbf{\textcolor{GoodMark}{\ding{51}}} & \textbf{\textcolor{GoodMark}{\ding{51}}} & \textbf{\textcolor{BadMark}{\ding{55}}} & \textbf{?} & \textbf{?} \\ [0.8em]
\tableline
\parbox[c]{5.2cm}{\centering (7) A Relativistic\\Pulsar Beam} & \textbf{\textcolor{GoodMark}{\ding{51}}} & \textbf{\textcolor{BadMark}{\ding{55}}} & \textbf{\textcolor{BadMark}{\ding{55}}} & \textbf{\textcolor{BadMark}{\ding{55}}} & \textbf{\textcolor{GoodMark}{\ding{51}}} & \textbf{\textcolor{GoodMark}{\ding{51}}} \\ [0.8em]
\tableline
\enddata
\vspace{0.8em}
\textbf{Reference:} (1) \cite{blandford1983, Cox1991}, (2) \cite{bychkov1975,marcelin1990}, (3) \cite{chevalier1975,sankrit1997,Smith2013,Blondin2017}, (4) \cite{kundt1983,veron1985,Porth2014}, (5) \cite{fesen1993,fesen1997,rudie2008, Black2015}, (6) \cite{morrison1985}, (7) \cite{shull1984,benford1984,michel1985,bietenholz1990}.
\end{deluxetable*}

Moreover, models (2), and (7) lack any strong directional preference. If they were responsible for shaping the northern jet, they would be expected to operate similarly in other regions where breakout features are observed, producing comparable structures in the breakout regions. However, the northern jet extends much farther beyond the nebula than any of the other breakouts, making these models less compelling for explaining the Crab's northern jet specifically. For model (3), it is unclear how the PWN breaks out of the ejecta in different directions. Variations in conditions, such as the density of the ejecta, will influence the outcome of the PWN breakout. In this scenario, the Crab's northern jet could arise from conditions different from those in other breakout regions.

Additional constraints come from the finding by \citet{rudie2008} that the jet and the nebula are roughly coeval. Models (4), (6), and (7) rely on processes that would develop later in the nebula's evolution. Although the timescales could be tuned, these scenarios do not naturally produce a structure present from the earliest phases.

Models (1), (2), and (7) naturally produce a cylindrical jet structure. However, the cylindrical morphology of the jet is not a necessary outcome of models (3), (5), or (6), which could generate extended structures of many shapes. Model (4) relies on density contrasts in the ISM to form the jet. When the ejecta expand into a low-density ISM region, their morphology is expected to become irregular, making it unlikely to maintain the cylindrical shape observed in the Crab's northern jet. The expansion velocity would also vary in different directions, resulting in non-ballistic behavior of the ejecta in the case of model (4).

Models (5) and (7) are based on a direct consequence of the ballistic expansion of the supernova ejecta. In contrast, models (1), (2), (3), and (6) suggest, but do not necessarily require, a different ejecta expansion rate along a specific direction of the remnant, which would account for the jet's extended length relative to the rest of the nebula. In these latter scenarios, the ejecta outside the jet direction are generally assumed to expand ballistically. 

Magnetic- or plasma-based models (2) and (7) also require strong magnetic fields that are generally assumed to emanate from the Crab pulsar. However, the optical jet is not aligned with either the pulsar's spin axis or with the plane of the PWN equatorial torus, which lies perpendicular to the spin axis. This misalignment presents a major difficulty for any model relying on pulsar-driven plasma instabilities \citep{rudie2008}. Model (6) faces the additional shortcoming that it is unable to explain the observed optical synchrotron emission from the jet \citep{veron1985,bietenholz1990}.

In summary, models (1), (3), and (5) do not exhibit strong inconsistencies with the available observations and therefore seem viable, whereas the remaining models show significant conflicts with one or more observed properties of the Crab. In the following section, we examine these three models in greater detail.

\subsection{Viable Models} 

\citet{blandford1983} first proposed model (1), the mass-loss trail scenario, in which the northern jet of the Crab Nebula arises from a low-density channel created by mass loss from the progenitor star during a red-giant phase prior to explosion. In this picture, the progenitor's motion through the interstellar medium would cause its stellar wind to be shocked and swept downstream, forming an extended trailing structure. Such a trail could evolve into an elongated, shell-like configuration, with its interior filled either by stellar wind material that was insufficiently shocked to collapse or by ambient interstellar gas. Analogous mass-loss-driven structures have been observed around other asymptotic giant branch stars and red supergiants, lending plausibility to this mechanism \citep{Cox2012AandA}.

However, the suggested small angle ($\approx 15^\circ$) to the LOS tilting of the trail from the model does not match the observed angle of the northern jet. \cite{shull1984} and \cite{marcelin1990} found an expansion speed for the jet away from its axis of 260 km s$^{-1}$, which is inconsistent with the lateral expansion velocity of less than 1 km s$^{-1}$ derived from the model \citep{Cox1991}. In addition, any radio emission from the trail would be thermal and, therefore, unpolarized, and would be much weaker than that observed by \cite{Wilson1985} and \cite{bietenholz1990}.

Alternatively, \cite{Cox1991} provided a revised version of the star mass-loss trail model to encounter these problems above. The trail suggested in the revised model lies in the plane of the sky outside the fast-moving ejecta. The interior of the trail tube is filled with gas from the stellar wind, which was not shocked strongly enough to collapse and is less dense than the ambient ISM. When the star explodes, some of the hot gas behind the supernova's bow shock travels up the tube. Since the interior of the tube is less dense than the ambient ISM, the hot gas travels up the tube faster than elsewhere, so the bow shock lies further from the explosion center inside the tube than outside \citep{Cox1991}. 

In order for the revised model to reproduce the observed properties of the northern jet, the velocity of the progenitor must be supersonic and greater than the velocity of its stellar wind, the Crab Nebula must be within the warm component of the ISM, the density of the mass-loss trail tube wall must be at least two orders of magnitude higher than that of the surrounding interstellar medium, and the outer \cite{Chevalier1977} halo must not exist \citep{Cox1991}. In addition to the highly restricted initial conditions for the revised model to work, there is no clear evidence for the recess of the nebula edge to either side of the jet, which the model predicts, caused by the interaction of the nebula's bow shock with the dense region around the bow shock of the star trail. 

Unlike the mass-loss trail hypothesis, model (5) has more direct observational evidence. The orientation of the synchrotron ``bays'' \citep{Fesen1992}, helium-rich filament populations near the center of the remnant \citep{MacAlpine1989,Fesen1992,Lawrence1995,Satterfield2012}, and direct kinematic measurements of the eject filaments \citep{fesen1993,fesen1997,rudie2008, Black2015} all suggest that there is N-S bipolar structure of the Crab nebula. 

Proper motion measurements of the Crab's northern jet ejecta suggested that the jet likely formed during the 1054 supernova explosion and represents the remnant's highest velocity knots \citep{rudie2008}. \cite{Black2015} found a large and nearly emission-free opening in the remnant's thick outer ejecta shell located directly below the Crab's northern jet and along the northernmost point of the Crab's N-S bipolar structure. The Crab's northern jet is located at the top of the bipolar expansion and over this opening \citep{Black2015}. Both proper motion measurement and location of the Crab's northern jet suggest that it is part of the Crab's N-S bipolar expansion structure.

Interpreting the Crab's northern jet as part of a larger north-south bipolar outflow naturally raises the question of whether a corresponding southern counter-jet should be present. To date, however, there is no clear observational evidence for such a feature. If a southern counter-jet does exist, it is clearly not as distinct or as morphologically prominent as the northern jet relative to the rest of the remnant. Interaction with the surrounding nebular environment may have distorted or deflected any southern counterpart, in a manner analogous to the deflection observed in the southern X-ray jet, potentially rendering it less collimated and less readily identifiable.

Relevant to model (3), proper-motion measurements have shown that the expansion of the PWN has displaced and accelerated the Crab's optical filaments relative to their expected positions under simple ballistic expansion \citep{Trimble1968,Wyckoff1977,bietenholz1990,Nugent1998}. As the PWN expands into the network of ejecta filaments, it may locally penetrate the thermal ejecta shell in regions where the ejecta density is relatively low. In Section 4.4, we identified several such breakout regions that may plausibly arise from this process. The Crab's northern jet could represent an extreme manifestation of one such breakout.

Despite this qualitative agreement, the physical details governing PWN penetration through the ejecta shell remain poorly understood. In particular, the geometry, stability, and penetration depth of such breakouts and the conditions under which they can be sustained are not well constrained. Producing a structure like the Crab's northern jet through this mechanism would likely require finely tuned initial conditions to generate both its well-defined, funnel-like morphology and its large radial extent beyond the main body of the nebula, distinguishing it from the smaller-scale breakout regions identified elsewhere in the remnant.

Each of the three viable models considered here offers a plausible explanation for certain observed properties of the Crab's northern jet, yet none alone fully accounts for its morphology, kinematics, and degree of collimation. Importantly, these scenarios are not mutually exclusive, and contributions from more than one process cannot be ruled out. However, the relative importance of each mechanism, and whether one dominates the jet's formation, remains unclear. Resolving this question will require fully three-dimensional hydrodynamic simulations that track the system's evolution from the progenitor mass-loss phase through the supernova explosion and subsequent interaction with the expanding pulsar wind nebula, while incorporating the observational constraints discussed above. Such simulations would also inform a broader understanding of how PWN expansion couples to asymmetric ejecta structure and progenitor mass loss history in other supernova remnants hosting PWN.

\section{CONCLUSIONS}

We have presented new three-dimensional kinematic reconstructions of the Crab Nebula using hyperspectral cubes from SITELLE, enabling a detailed investigation of the structure, geometry, and origin of the northern jet. Using [\ion{O}{3}] $\lambda$$\lambda$4959, 5007 emission, we obtain the most comprehensive view to date of the jet's kinematics and its interface with the surrounding ejecta. Our principal findings are as follows:

1) The Crab's northern jet extends approximately 0.6 pc in the radial direction and 0.4 - 0.5 pc east-west, with funnel walls of thickness roughly 0.05 - 0.1 pc. The jet is inclined by about $8 \pm2$ degrees into the plane of the sky and by $5 \pm 2$ degrees toward the west. These inclination angles and the jet's elliptical, segmented funnel morphology are consistent with earlier observational studies. The integrated spectral study of the emission within the jet funnel region demonstrates that emission is present inside the Crab northern jet, a region previously considered empty or unclear on the basis of earlier observations. No significant differences are found between the emission within the funnel and that in the jet wall.

2)  Below the jet's base lies a nearly circular cavity, located about 0.9 pc from the expansion center and exhibiting a radius of 0.3 - 0.4 pc with a thin wall of order 0.03 pc. The close similarity in size, spatial alignment, and inclination between the cavity and the jet strongly indicates that they are physically related components of a single structural system.

3) Four regions where filaments of emission extend conspicuously beyond the main cage of the nebula are clearly identified in [\ion{O}{3}] $\lambda$$\lambda$4959, 5007 emission. These breakouts differ in both length and morphology, yet each aligns with outward extensions of the synchortron nebula. Their geometry strongly suggests that the expanding synchrotron bubble has displaced nearby ejecta, pushing material beyond the [\ion{O}{3}] emitting shell and producing the observed breakout structures.

Taken together, these observations point to several constraints on the formation of the Crab's northern jet. The jet's morphology and kinematics strongly suggest that the early PWN played a central role in its development, rather than the explosion geometry alone. The ballistic velocity profile of the jet ejecta \citep{rudie2008}, implies that the period of significant PWN influence was short relative to the dynamical age of the remnant, since extended momentum input would have produced higher ejecta velocities than those observed. In addition, the jet's long, coherent structure requires that ejecta along this axis were not strongly affected by Rayleigh-Taylor or Kelvin-Helmholtz instabilities. A brief but energetic early PWN interaction provides a mechanism for imprinting a stable velocity field before such instabilities have time to grow.

Several formation scenarios remain consistent with these constraints. One possibility is that the jet represents the highest-velocity component of a modest N-S bipolar expansion, potentially enhanced by the pinching effect of a He-rich circumstellar disk \citep{Black2015}. This mechanism could produce an axis of fastest expansion, but it does not readily explain why the northern extension is significantly more collimated than other observed breakout features, nor why a southern counterpart is not present.

Following \citet{Smith2013}, a second possibility is that a breach or localized underdensity in the ejecta shell provided a low-resistance corridor through which the early PWN preferentially expanded. Although such a feature could facilitate channeling, a simple break or pocket in the shell would likely be susceptible to Rayleigh-Taylor and Kelvin-Helmholtz instabilities that would erode its walls and compromise collimation. A sufficiently strong and brief PWN episode might stabilize such a channel, but the required physical conditions are not yet well constrained.

A third scenario invokes a pre-existing mass-loss trail from the progenitor, which could establish a lower-density pathway through the ejecta that acts as a waveguide for PWN-accelerated material \citep{blandford1983,Cox1991}. This model provides a natural mechanism for channeling and collimation, but it also faces uncertainties. Most notably, is whether the necessary density contrast and spatial coherence could survive through core collapse and early expansion. In addition, if the jet walls are indeed high-velocity ejecta rather than remnants of a pre-existing tunnel, it becomes unclear whether this scenario can account for their observed thinness and ∼$\sim$2000 km\,s$^{-1}$ proper motions.

At present, none of these scenarios can be distinguished on the basis of current observations alone. Moreover, the proposed mechanisms are not necessarily mutually exclusive, and contributions from more than one process may be involved, although the extent of any such overlap remains uncertain. Further progress will require three-dimensional hydrodynamic simulations that explore a range of PWN energy-injection timescales, ejecta-shell geometries, and circumstellar density distributions, and that can be directly compared with the observed morphology and kinematics of the jet. The proper-motion age of the jet ejecta \citep{rudie2008}, provides a useful upper limit on the duration and magnitude of early PWN influence and should serve as an important constraint for such models. Finally, refining the distance to the Crab, particularly in light of evidence that a value closer to 1.6 kpc would reduce the inferred physical extent in the plane of the sky and  alleviate the apparent compression of the nebula along the LOS (radial-velocity) dimension, will be essential for establishing accurate three-dimensional reconstructions and for ensuring that future simulations are compared against physically consistent spatial scales.

\begin{acknowledgments}
We thank an anonymous referee for insightful comments that helped improve the content and presentation of this paper. D.M.\ acknowledges support from the National Science Foundation through grants PHY-2209451 and AST-2206532. T.T. acknowledges support from the NSF grant AST-2205314 and the NASA ADAP award 80NSSC23K1130. We are grateful to Robert Fesen and Nathan Smith for valuable discussions that enriched the scientific context and breadth of this paper. Maxim Lyutikov, William Blair, and Ravi Sankrit provided helpful comments on an earlier draft of the manuscript.
\end{acknowledgments}

\appendix

\renewcommand\thefigure{A\arabic{figure}}    
\setcounter{figure}{0}   

\begin{figure}
\begin{subfigure}{.5\textwidth}
  \centering
  \includegraphics[width=1\linewidth]{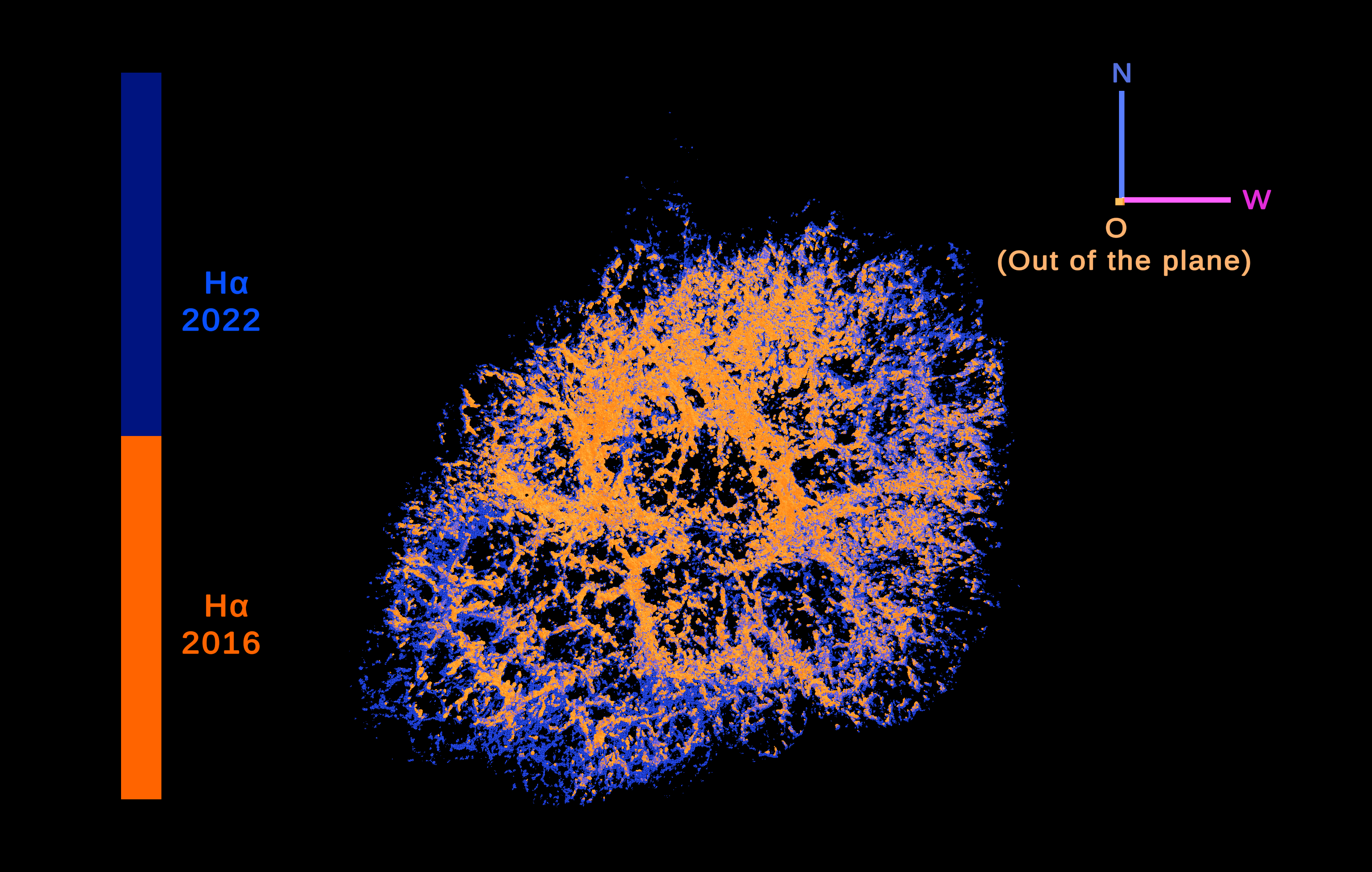}  
\end{subfigure}
\begin{subfigure}{.5\textwidth}
  \centering
  \includegraphics[width=1\linewidth]{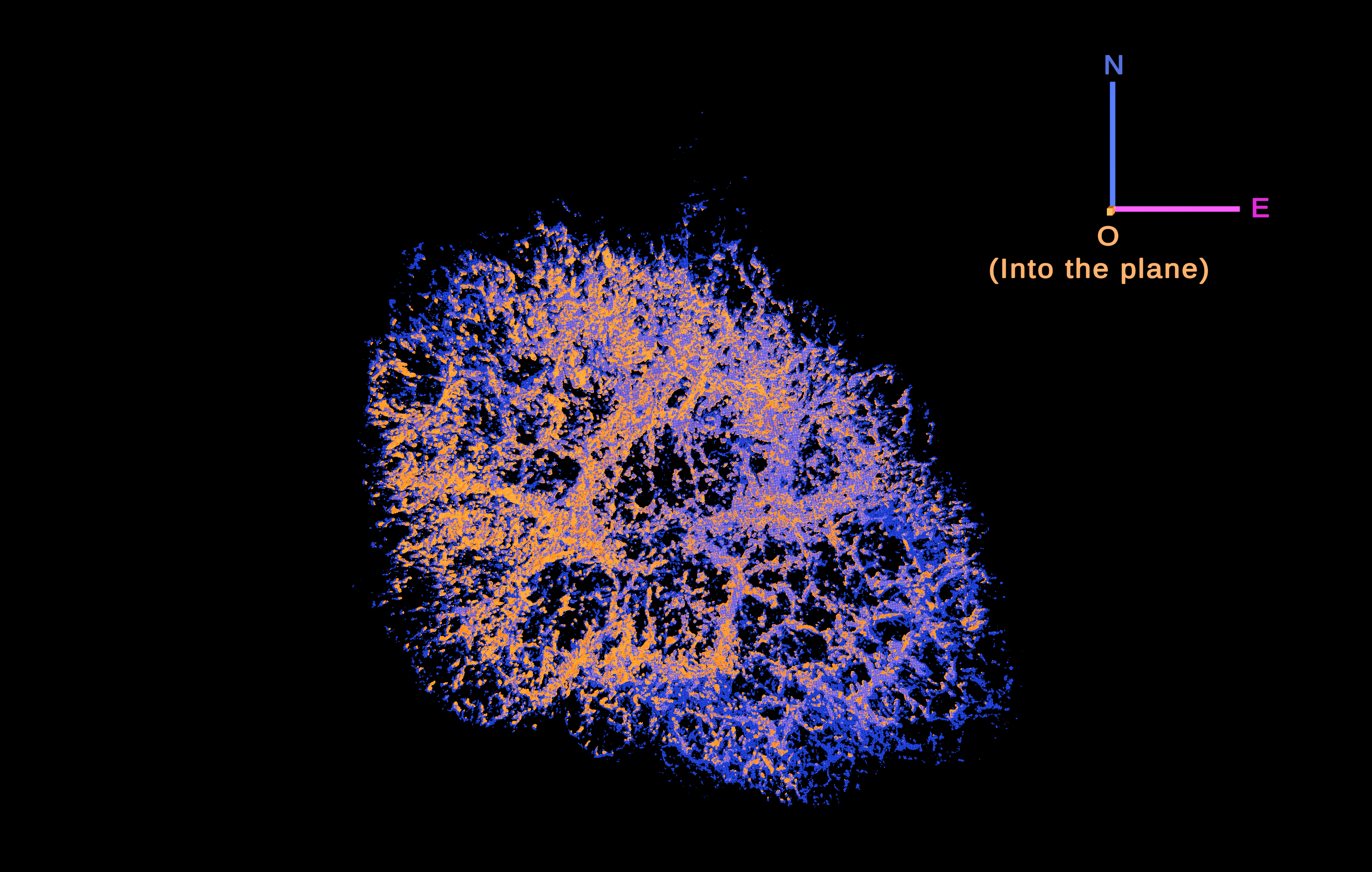}  
\end{subfigure}
\captionsetup{font=small}
\caption{Comparison between our new 3-D reconstructions result (blue) in H$\alpha$ emission and the previous 3-D reconstructions result (orange) in H$\alpha$ emission from 2016 SITELLE observations \citep{Martin2021}. The expansion of the nebula over the six-year interval between the two datasets is visible in the comparison.}
\label{fig:oldnew_ha}
\end{figure}

\begin{figure}
\begin{subfigure}{.5\textwidth}
  \centering
  \includegraphics[width=1\linewidth]{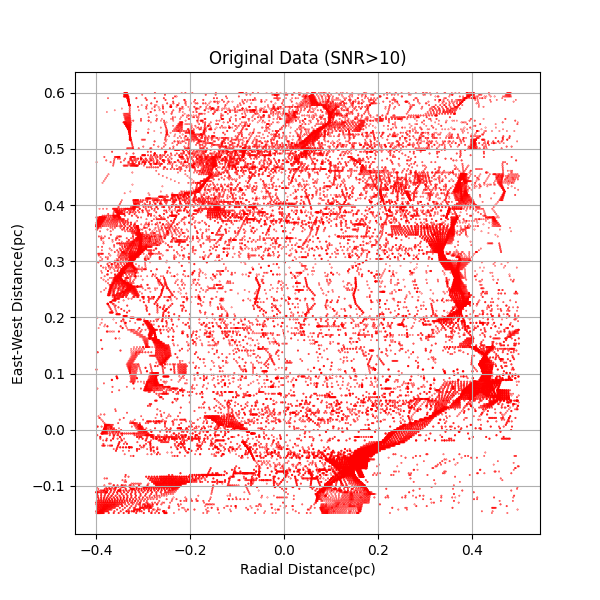}  
  \caption{Original Data}
\end{subfigure}
\begin{subfigure}{.5\textwidth}
  \centering
  \includegraphics[width=1\linewidth]{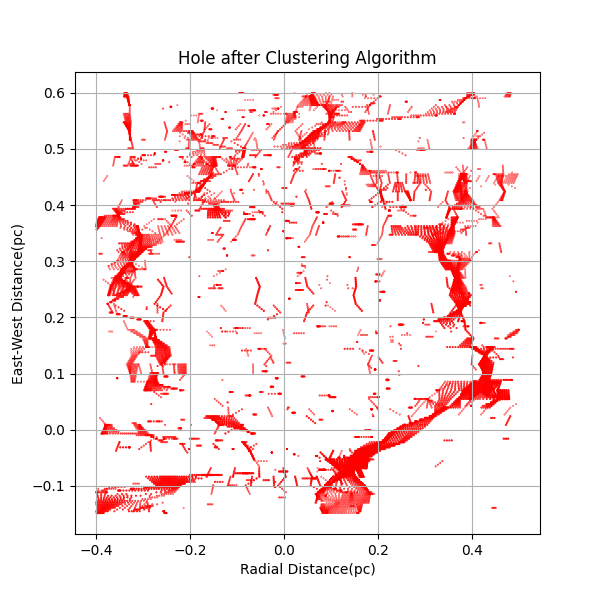}  
  \caption{Clustering Method}
\end{subfigure}
\begin{subfigure}{.5\textwidth}
  \centering
  \includegraphics[width=1\linewidth]{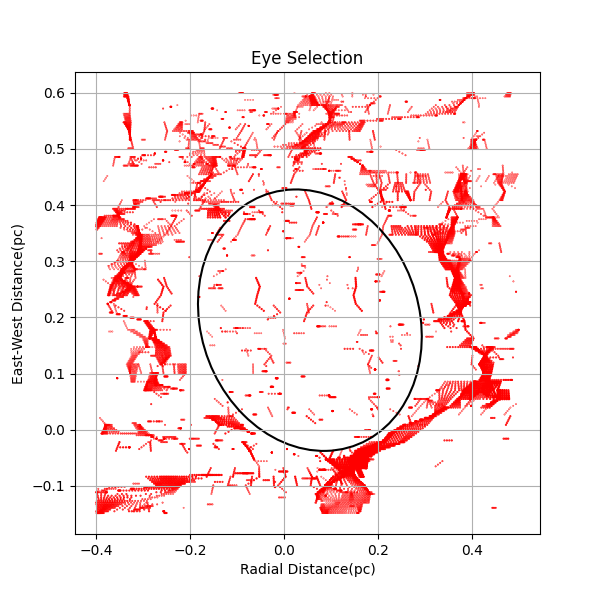}  
  \caption{Eye Select}
\end{subfigure}
\begin{subfigure}{.5\textwidth}
  \centering
  \includegraphics[width=1\linewidth]{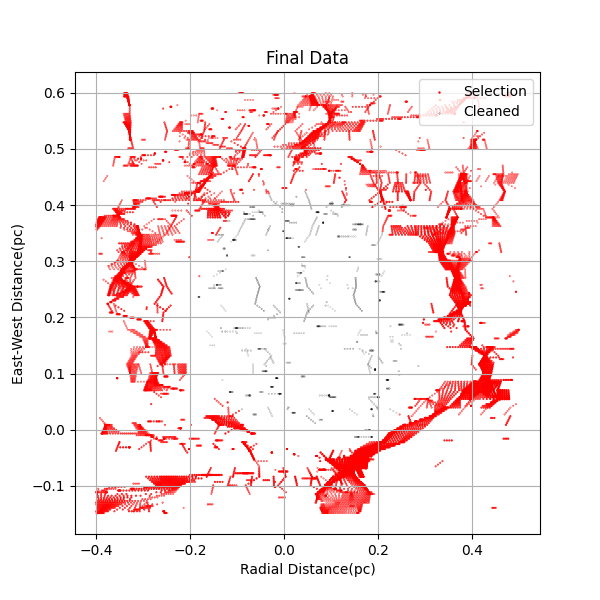}  
  \caption{Final Data Set}
\end{subfigure}
\caption{Hole region data processing. A step-by-step demonstration of the data reduction process applied to the [\ion{O}{3}] $\lambda\lambda$4959, 5007 emission, showing how the annular structure becomes best represented in the hole region. The plots are shown viewed from the top of the hole/nebula. The size of all the data points represent the flux ratio compared to the highest flux value in the entire jet region.}
\label{fig:hole4}
\end{figure}

In Figure~\ref{fig:oldnew_ha}, we compare our new H$\alpha$ 3D reconstruction result with the H$\alpha$ result from the previous work of \cite{Martin2021}, which was based on SITELLE observations obtained in 2016. Our new reconstruction resolves the nebula in significantly greater detail, an improvement of approximately 80$\%$, with 755,912 data points compared to 419,044 in the previous reconstruction. The expansion of the nebula over the six-year interval between the two datasets is also visible when the reconstructions are compared.

In Figure~\ref{fig:hole4}, we illustrate all the steps taken to process the data in the hole region, which result in the color-coded data used to produce Figure~\ref{fig:hole_all} for the [\ion{O}{3}] $\lambda\lambda$4959, 5007 emission. The size of each data point represents its flux relative to the highest flux value in the entire hole region, with larger points corresponding to higher flux values.

\newpage
\bibliography{Reference}{}
\bibliographystyle{aasjournal}

\end{CJK*}
\end{document}